\documentclass[a4paper,aps,prd,showpacs,nofootinbib,preprintnumbers,amsmath,amssymb,mcite,12pt]{revtex4-1}

\usepackage{amsfonts,amsmath,amssymb,amsthm}
\usepackage{axodraw4j,pstricks,color}
\usepackage{latexsym}
\usepackage{graphicx}
\usepackage{slashed}
\usepackage[colorlinks]{hyperref}
\hypersetup{colorlinks=true,backref=true,linkcolor=black,anchorcolor=black,citecolor=black,filecolor=black,menucolor=black,pagecolor=black,urlcolor=black}

\numberwithin{equation}{section}

\def\stamp{--- {\bf \today} --- {\bf \jobname.tex}}

\def\tr{\textrm{tr}}

\def\cN{\mathcal{N}}
\def\cP{\mathcal{P}}

\def\stamp{--- {\bf \today} --- {\bf \jobname.tex}}

\def\tr{\textrm{tr}}

\def\tr{\textrm{tr}}

\def\nn{\nonumber}
\def\sign(#1){\textrm{sign}(#1)}

\def\cP{\mathcal{P}}

\def\cN{\mathcal{N}}

\def\BE{\begin{equation}}
\def\EE{\end{equation}}

\def\sla#1{\not\!{#1}}

 \def\<#1|#2){\left\langle#1|#2\right\rangle}
 \def\<#1|#2|#3]{\left\langle#1|#2|#3\right ]}
\def\(#1|#2|#3>{\left[#1|#2|#3\right\rangle}
 \def\[#1|#2]{\left[#1|#2\right]}

\def\an[#1,#2]{\left\langle#1\,#2\right\rangle}
\def\aq[#1,#2,#3]{\left\langle#1|#2|#3\right]}
\def\qa[#1,#2,#3]{\left[#1|#2|#3\right\rangle}
\def\sq[#1,#2]{\left[#1\,#2\right]}
\def\spa#1.#2{\left\langle#1\,#2\right\rangle}
\def\spab[#1,#2,#3]{\left\langle#1|#2|#3\right]}
\def\spba[#1,#2,#3]{\left[#1|#2|#3\right\rangle}
\def\spb#1.#2{\left[#1\,#2\right]}
\def\lor#1.#2{\left(#1\,#2\right)}

\begin{document}

%%%%%%%%%%%%%%%%%%%%%%%%%%%%%%%%%%%%%%%%%%%%%%%%%%%%%%%%%%%%%%%%%
\preprint{IHES/P/13/23, IPHT-t13/019, ACFI-T13-03}
\title{On-shell Techniques and Universal Results in Quantum Gravity}
\author{\vspace{.5cm}\bf N.~E.~J~Bjerrum-Bohr${}^a$, John~F.~Donoghue${}^{a,b}$ and Pierre~Vanhove${}^{a,c,d}$}
\affiliation{${}^a$  Niels Bohr International Academy and Discovery Center,\\
The Niels Bohr Institute, Blegdamsvej 17,\\
DK-2100 Copenhagen \O, Denmark,\\
${}^b$Department of Physics, University of Massachusetts,
Amherst, MA 01003, USA\\
${}^c$ CEA, DSM, Institut de Physique Th{\'e}orique, IPhT, CNRS, MPPU,\\
URA2306, Saclay, F-91191 Gif-sur-Yvette, France\\
${}^d$ Institut des Hautes {\'E}tudes Scientifiques
Bures sur Yvette, F-91440, France
}
\email{bjbohr@nbi.dk, donoghue@physics.umass.edu, pierre.vanhove@cea.fr}

\begin{abstract}
\vspace{0.7cm}
We compute the leading post-Newtonian and quantum corrections
to the Coulomb and Newtonian potentials using the full modern arsenal
of on-shell techniques; we employ spinor-helicity variables everywhere,
use the Kawai-Lewellen-Tye (KLT) relations to derive gravity amplitudes from gauge
theory and use unitarity methods to extract the terms needed at one-loop order.
We stress that our results are {\it universal} and thus will hold in any quantum
theory of gravity with the same low-energy degrees of freedom as we are considering.
Previous results for the corrections to the same potentials, derived historically using
Feynman graphs, are verified explicitly, but our approach presents a huge
simplification, since starting points for the computations are compact and
tedious index contractions and various complicated integral reductions
are eliminated from the onset, streamlining the derivations.
We also analyze the spin dependence of the results using
the KLT factorization, and show how the spinless corrections in the framework
are easily seen to be independent of the interacting matter considered.
\end{abstract}

\maketitle
\setcounter{tocdepth}{0}
\tableofcontents

%%%%%%%%%%%%%%%%%%%%%%%%%%%%%%%%%%%%%%%%%%%%%%%%%%%%%%%%%%%%%%%%%
\section{Introduction}\label{sec:introduction}
Unitarity based methods combined with the helicity formalism have proven exceptionally
successful in gauge theory calculations at one loop (see {\it e.g.}~\cite{Dixon:1996wi, Ellis:2011cr}).
Such methods have so far been less frequently applied to
general relativity~\cite{Bern:2002kj, Dunbar:1995ed, Neill:2013wsa}, and
quantum corrections to gravitational systems with massive matter have not
been studied in this framework at all. However, such techniques are well-suited
for effective field theory considerations in low energy quantum gravity. Early
treatments of gravitational loops tended to focus on the ultraviolet divergences,
but effective field theory methods have allowed us to separate these
ultraviolet divergences from the universal reliable predictions of the low energy portion
of the theory~\cite{Donoghue:dn,BjerrumBohr:2002ks,BjerrumBohr:2002kt, Khriplovich:2002bt,Burgess:2003jk,Holstein:2008sx,Donoghue:2012zc}.
The unitarity methods deal directly with on-shell and low energy amplitudes, and
products of on-shell tree amplitudes can therefore yield the low energy one-loop
results in a conceptually simple manner.

In this paper, we apply new on-shell amplitude methods to the gravitational scattering
of massive matter~\cite{Donoghue:dn, BjerrumBohr:2002ks,BjerrumBohr:2002kt,
Khriplovich:2002bt}, our focus here is especially on the quantum corrections to
the potential (The ref.~\cite{Neill:2013wsa}
which appeared near the end of our work, also
uses on-shell methods to derive corrections to the scattering potential,
but there only the classical correction terms are considered).
The unitarity cut for the leading quantum corrections involves the gravitational
Compton amplitude, {\it i.e.} the two on-shell gravitons coupled
to matter. For matter fields of all spins, this amplitude has a simple structure,
as it is related to the square of the electromagnetic Compton amplitude (involving
photons)~\cite{Choi:1993wu,Choi:1993xa,Holstein:2006pq,Holstein:2006bh,Bern:2002kj}.
A useful observation for our calculation is that computing the massless two-particle cut
gives us exactly everything we need. The cut of the amplitude is
precisely one-to-one with the non-analytic parts of the amplitude that contributes
to the long-distance leading corrections to the scattering potential
at one-loop. Hence, we do not need to reconstruct the full amplitude - we
only need to consider the terms contributing to the massless two-particle cut.

Moreover, there is an added bonus in using the cut and decomposing the amplitudes
using KLT; in such a setup one can easily dissect the interaction
between the two particles into a series of spin corrections; {\it i.e.} a coefficient
for the spinless interaction and coefficients of spin-spin interactions, {\it etc.}
It has been seen before in direct calculations~\cite{Holstein:2008sx} 
that the coefficients themselves are actually independent of the type
of interacting matter; that is, they are the same for massive scalars,
fermions or vector bosons. We call this property {\sl matter universality}
or {\sl matter independence}.
This observation appears however to be somewhat puzzling in the context of Feynman diagrams, because here
the vertex rules (and even the diagrams that need to be calculated) differ greatly
for different types of matter particles. In this paper our focus will be on the spinless interaction
part of the series of spin corrections. We will demonstrate directly using the on-shell cut method and KLT
that this coupling is always identical for any type of particle interaction; i.e. in the non-relativistic limit
it only depends on the masses of the interacting particles.

The classical and quantum corrections to the Newtonian potential can be
addressed by studying the scattering matrix element in the non-relativistic
limit
\begin{equation}\label{e:MqDef}
\langle p_1,p_2| i T|p_3,p_4\rangle= -i \,M(q)\, (2\pi)^4\,
\delta^{(4)}(\sum_{i=1}^4 p_i)\,,
  \end{equation}
where $p_i$ with $i=1,2,3,4$ is the incoming momentum of the individual particles and   $q= p_1+ p_2$ is the momentum transfer.
In momentum space (in the non-relativistic and free particle limit) we employ
the following definition of the scattering
potential $V(q)$ from the  amplitude
\begin{equation}\label{e:Vqdef}
  V(q) = {M(q)\over 4m_1m_2}\,,
\end{equation}
which provides a useful gauge invariant amplitude with which to display the quantum corrections~\cite{BjerrumBohr:2002kt}. One could, for example, have subtracted off the second order Born
contribution, which would lead to a nonrelativistic potential of the type used
in bound state quantum mechanics. We will not do this in this work.
The one-loop diagrams produce modifications to the tree interaction leading to a potential of the form
 \begin{equation}\label{e:Vq}
 V(q) =  {G_N m_1m_2\over {\vec q}^{\;2}}\left[-4\pi+C^{NP} \,G_N{(m_1+m_2)  \sqrt{ |\vec q|^2}}+
 G_N\hbar\, {\vec q}^{\;2}\left(C^{QG}\log({\vec q}^{\;2})+ \tilde C^{QG} \right)\right]\,.
\end{equation}
If this object is Fourier transformed to form a spatial potential, the term with the square-root
yields the classical $G_Nm/r$ general relativistic correction to the potential, and the term
with the logarithm produces a long-distance $G_N\hbar /r^2$ quantum correction.
The analytic correction without a logarithm will yield a short range $\delta^3 (r)$ effect in
the potential. The non-analytic terms (the square-root and the logarithm) arise from
long-distance propagation of the massless gravitons, and hence are genuinely
low-energy quantum predictions. These can be calculated in the effective field
theory approach. The analytic correction $\tilde C^{QG}$, however is not a prediction
of the low-energy theory as it is sensitive to the coefficients of higher curvature terms
in the gravitational action.

Our work in the present paper will focus on the square-root and logarithmic non-analytic
terms of the scattering potential.

The plan of the paper is as follows. In Sec.~\ref{sec:treeamp} we discuss the relations between
the gravitational Compton amplitude to the square of the electromagnetic one.
In Sec.~\ref{sec:helicity} we compute the one-loop amplitude in the helicity
formalism. Here we first calculate the electromagnetic case as a
warm up before moving on to our primary interest of the gravitational interaction. In
Sec.~\ref{sec:harmonic} we evaluate the amplitude in the covariant
harmonic gauge and compare with the Feynman approach used in earlier
computations.
In Sec.~\ref{sec:matter-independence} we discuss the matter-independence of the
non-analytic long-range contributions to the amplitude.
Finally,  Sec.~\ref{sec:conclusion} contains our conclusions and discussion. In
Appendix~\ref{sec:vertices} we list the covariant Feynman rules and
Appendix~\ref{sec:dispersion} discusses an alternative  evaluation of the cut
using dispersion relations.

%-------------------------------------------------------------------------

\section{The gravitational Compton amplitude}
\label{sec:treeamp}
In this section we will show how one can represent the gravitational
Compton scattering of two gravitons off a massive target of
spin~$s=0,\frac12, 1$ as the square of the QED (Abelian) Compton
scattering. We will  do this first using covariant amplitudes, and then
more compactly using the helicity formalism. The advantage of this approach
is that one can use the known expressions for the massive tree-level
amplitudes in Yang-Mills and QED to obtain in a condensed way
the massive tree-level amplitudes in gravity. As well, the connection
between the gravity and the QED amplitude will be instrumental in deriving
the matter-independence results in section~\ref{sec:matter-independence}.

%-----------------------------------------------------------------------
\subsection{Covariant notation}
\label{sec:gravcompton}
We will evaluate the one-loop amplitude by considering the
unitarity cut across the graviton lines in
section~\ref{sec:grav-compt-scatt} and~\ref{sec:oneloophel}, thus we need
to construct the tree-level amplitudes for the emission of two gravitons.

The tree amplitudes needed in this analysis can be constructed in
various way.  One direct covariant approach is to use the background
field vertices derived in~\cite{BjerrumBohr:2002ks,BjerrumBohr:2002kt}.
These vertices are listed in Appendix~\ref{sec:vertices}.  The vertex
$\tau_1^{\mu\nu}(p_1,p_2)$ given in eq.~\eqref{e:tau1} describes
the emission of a graviton from a massive scalar exchange.  Because the
metric is realized through the stress-energy tensor, the vertex couples
identically to quantum $h^{\mu\nu}$ or background
fields $H^{\mu\nu}$ (as used in refs.~\cite{BjerrumBohr:2002ks,BjerrumBohr:2002kt}).
The vertex $\tau_2^{\mu\nu;\rho\sigma}(p_1,p_2)$ given
in eq.~\eqref{e:tau2} is the four point interaction between two massive
scalars and two gravitons. Again the coupling between gravity and the
scalar through the stress-energy tensor implies that these vertices are
the same for quantum or background fields.
\begin{figure}[ht]
  \centering
  \includegraphics[width=3.5cm]{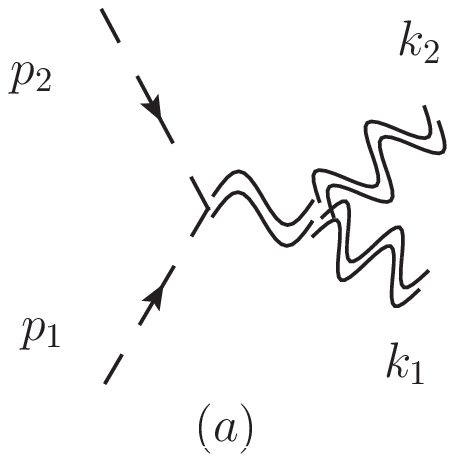}
  \includegraphics[width=3.5cm]{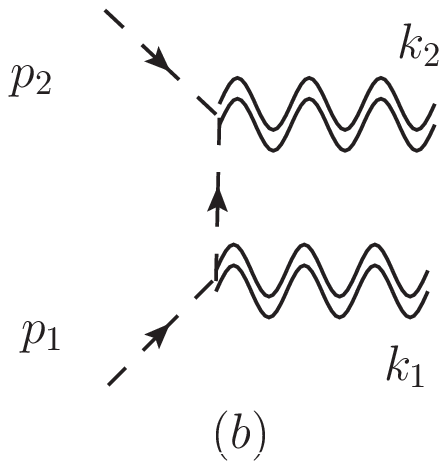}
  \includegraphics[width=3.5cm]{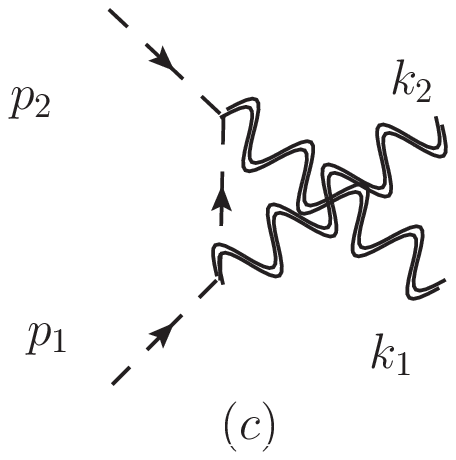}
  \includegraphics[width=3.5cm]{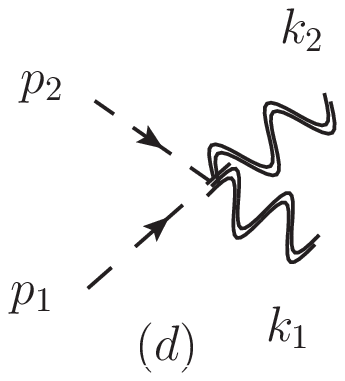}
  \caption{The various contributions to the tree-amplitude $\phi +
    \phi\to 2~gravitons$: (a) $s$-channel, (b) $t$-channel, (c)
    $u$-channel, (d) contact term.}
  \label{fig:treeemission}
\end{figure}

In order to compute the general relativity correction and the
quantum correction arising from the one-loop diagram we need the
tree-level amplitude for emitting two gravitons as illustrated in figure~\ref{fig:treeemission}.
In the covariant  approach using the background field vertices the
tree-level amplitude for emitting two quantum gravitons $h$ of
polarization $\epsilon_1^{\alpha\beta}(k_1)$ and $\epsilon_2^{\gamma\delta}(k_2)$ is given
by (with all incoming  external momenta)
\begin{eqnarray}
  \label{e:treeCov1}
i  M^{\rm tree}(p_1,p_2,k_1,k_2)&=&\tau_1^{\mu\nu}(p_1,p_2) \, {i
    \cP_{\mu\nu;\rho\sigma}\over q^2+i\varepsilon }\,
  \tau_3^{\rho\sigma}{}_{\alpha\beta;\gamma\delta}(k_1,k_2,p_1+p_2)\,
  \epsilon_1^{\alpha\beta}(k_1)\epsilon_2^{\gamma\delta}(k_2)\cr
&+&{\tau_{1\,\alpha\beta}(p_1,-p_1-k_1) \, i \,\tau_{1\,\gamma\delta}(p_1+k_1,p_2)
    \over (p_1+k_1)^2-m^2+i\varepsilon }
  \, \epsilon_1^{\alpha\beta}(k_1)\epsilon_2^{\gamma\delta}(k_2)\cr
&+&{\tau_{1\,\gamma\delta}(p_1,-p_1-k_2) \, i \,\tau_{1\,\alpha\beta}(p_1+k_2,p_2)
    \over (p_1+k_2)^2-m^2+i\varepsilon }
  \, \epsilon_1^{\alpha\beta}(k_1)\epsilon_2^{\gamma\delta}(k_2)\cr
&+&\tau_{2\,\alpha\beta;\gamma\delta}(p_1,p_2) \,
\epsilon_1^{\alpha\beta}(k_1)\epsilon_2^{\gamma\delta}(k_2)\,,
\end{eqnarray}
with
\begin{equation}\label{e:defP}
\cP_{\alpha\beta;\gamma\delta} =
\frac12\left[\eta_{\alpha\gamma}\eta_{\beta\delta} +
\eta_{\beta\gamma}\eta_{\alpha\delta}
-\eta_{\alpha\beta}\eta_{\gamma\delta}\right]\,,
\end{equation}
in  harmonic or de Donder gauge~\cite{Veltman:1975vx}. The three-graviton vertex
$\tau_3^{\mu\nu}{}_{\alpha\beta;\gamma\delta}$, given in eq.~\eqref{e:tau3},
between two quantum
fields $h$ and one background field $H$ differs from the vertex
for three quantum gravitons derived by De Witt~\cite{Dewitt} and
Sannan~\cite{Sannan:1986tz}.
We have checked that the on-shell amplitude constructed with the
three-graviton vertices derived in~\cite{Dewitt,Sannan:1986tz} leads to the
same answer as ours.
Notice that its expression given in~\eqref{e:tau3} is  much
simpler  than the three-graviton vertex of these references.

We have also checked that our amplitude correctly satisfies the
relation to the QED amplitude~\cite{Choi:1994ax} which we discuss below
in the context of the helicity formalism.

%-----------------------------------------------------------------------
\subsection{Massive trees  amplitude in gravity from Yang-Mills tree amplitudes}
\label{sec:gravity-as-square}
A different approach is to construct the gravity amplitudes by
applying the KLT method to the emission of two gluons from massive scalars.

The KLT relation between massless four-point gravity amplitudes and Yang-Mills
amplitudes reads~\cite{Kawai:1985xq,Bern:2008qj}
\begin{equation}\label{e:Grav1}
i  M_s^{\rm tree} (p_1,p_2,k_1,k_2)=\kappa_{(4)}^2\,  ( p_1\cdot
k_1)\, A_s^{\rm tree}(p_1,p_2,k_2,k_1) A_0^{\rm tree}(p_1,k_2,p_2,k_1)\,.
\end{equation}
Where $M_s^{\rm tree}(p_1,p_2,k_1,k_2)$ is the tree-level scattering,
$X^s(p_1) h(k_1)\to X^s(-p_2) h(-k_2)$ with $p_1+p_2+k_1+k_2=0$, 
between a matter field $X^s$ of spin $s=0,\frac12,1$ and gravitons
$h$, given by the sum of diagrams in fig.~\ref{fig:treeemission}.  We use
$\kappa_{(4)}^2=32\pi G_N$.  The gauge theory amplitude $A_s^{\rm
  tree}(p_1,p_2,k_2,k_1)$ is the tree-level scattering amplitude
between a matter field $\phi^s$ of spin $s=0,\frac12,1$ and gluons
$\phi^s(p_1) (\phi^s(p_2))^*\to g(-k_1)g(-k_2)$. The amplitude
$A_0^{\rm tree}(p_1,k_2,p_2,k_1)$ is the tree-level scattering between
a scalar matter field $\phi^0$ and gluons $\phi^0(p_1)g(k_1)\to
\phi^0(-p_2) g(-k_2)$.

The color-stripped ordered Yang-Mills amplitudes satisfy the
amplitude relation~\cite{Bern:2008qj}
\begin{equation}\label{e:mon}
     A_s^{\rm tree}(p_1,p_2,k_2,k_1)={p_1\cdot k_2\over k_1\cdot k_2}\,  A_s^{\rm tree}(p_1,k_2,p_2,k_1)\,,
\end{equation}
allowing us to express the amplitude in~\eqref{e:Grav1} in the following manner,
\begin{equation}\label{e:Grav2}
i  M_s^{\rm tree} (p_1,p_2,k_1,k_2)={\kappa_{(4)}^2\over e^2}  { (p_1\cdot
k_1)\, p_1\cdot k_2\over k_1\cdot k_2}\, A_s^{\rm tree}(p_1,k_2,p_2,k_1) A_0^{\rm tree}(p_1,k_2,p_2,k_1)\,.
\end{equation}

We will now explain that these amplitude relations are valid in the same form replacing massless fields  with massive matter fields. The general
form of these massless amplitudes
for $n$-point color-ordered  gauge theory amplitudes
$A_n^{\rm tree}(\sigma)$ and the $n$-point gravity amplitudes
$M_n^{\rm tree}$ takes the form~\cite{Kawai:1985xq,Bern:2008qj,BjerrumBohr:2010ta,BjerrumBohr:2010hn}
\begin{multline}
  iM^{\rm tree} =\sum_{\sigma, \gamma\in\mathfrak S_{n-3}} \mathcal S[\sigma(2,\cdots,n-2)|\gamma(2,\cdots,n-2)]|_{k_1}\times
\cr
\label{e:KLT}\times  A^{\rm tree}(1,\sigma(2,\cdots,n-2),n-1,n) A^{\rm
    tree}(n,n-1,\gamma(2,\cdots,n-2),1)  \,.
\end{multline}
with $\mathfrak S_{n-3}$ denoting the possible permutations over $n-3$ indices and where
 the momentum kernel $\cal S$ is given by the expression
\begin{equation}
  \label{e:Skernel}
  S[i_1,\dots, i_r|j_1,\dots, j_r]|_p=\prod_{t=1}^r (p\cdot
  k_{i_r}+\sum_{s>t}^r \theta(i_r,i_s)\, k_{i_r}\cdot k_{i_s} )\,.
\end{equation}
Here $\theta(i_t,i_s)$ equals 1 if the ordering of the legs $i_r$ and $i_s$
is opposite in the sets $\{i_1,\dots,i_r\}$ and $\{j_1,\dots,j_r\}$, and 0 if the
ordering is the same.

This relation can be rewritten in various equivalent way thanks to the
BCJ relations~\cite{BCJ} or more conveniently for our purposes in terms of the momentum
kernel~\cite{BjerrumBohr:2009rd,BjerrumBohr:2010hn}
\begin{equation}
  \sum_{n\in\mathfrak S_{n-2}} \,  \mathcal S[\sigma(2,\cdots,n-1)|\gamma(2,\cdots,n-2)]|_{k_1}
\times  A^{\rm tree}(1,\sigma(2,\cdots,n-1),n)=0; \qquad
\forall\gamma\in\mathfrak S_{n-2}\,,
\end{equation}
generalizing the relation in eq.~\eqref{e:mon}.

These relations have been derived for a number of different types of
matter including, massless scalars, vectors (gluons or photons), and
gravitons~\cite{Bern:1999bx}. The derivation shows that the relation
is the same in any space-time dimensions. However, the key point is that a
massive scalar in four dimensions is equivalent to a massless scalar
with momenta living in higher dimensions.
Therefore, an amplitude between massive scalars
and gravitons in four dimensions, can be seen as a tree-level
amplitude between massless scalars in higher dimensions with gravitons
polarized in four-dimensions. In this higher-dimensional setup the
relation between gravity and gauge theory can be applied.

The validity of the amplitude relations with
massive scalars and gravitons also follows directly from
string theory. The case of tachyons was already considered
in~\cite{BjerrumBohr:2010zs}.
The relations in~\cite{Kawai:1985xq,BjerrumBohr:2010hn} relies on
the monodromy properties of the colored-ordered open string amplitudes
\begin{equation}
A_{\alpha'} (i_1,\dots,i_n)= \int_{x_{i_1}<\cdots <x_{i_n}}
\, f(x_i-x_j)  \, \prod_{i,j=1\atop
i\neq j}^n \, (x_i-x_j)^{2\alpha' k_i\cdot k_j}\, \prod_{i=1}^n dx_i\,.
\end{equation}
The monodromy property however does not depend on the detailed expression for the
function $f(x_i-x_j)$ and is derived
from momentum conservation $\sum_i k_i=0$ and the phase factor that arises when going
around the branch cut given the factors $(x_i-x_j)^{2\alpha' k_i\cdot
k_j}$. The phase factor is not affected by any integer shifts of $2\alpha' k_i\cdot
k_j$ arising {\it e.g.} from a pole from
a massive scalar in the function
$f(x_i-x_j)$.
Thus the massive field theory amplitude relations
obtained by considering the $\alpha'\to0$ limit, satisfy the same
properties as explained in~\cite{BjerrumBohr:2010hn} as corresponding massless ones.

Assured that the KLT relation applies for various
types of matter fields, massless and massive, we will now study the case of four-point
amplitudes describing the emission of two gravitons from a matter field
of spin $s$.

%-----------------------------------------------------------------------
\subsection{Application of KLT to the gravitational Compton scattering:\\ Reduction to QED amplitudes}
\label{sec:grav-compt-scatt}
\begin{figure}[ht]
  \centering
  \includegraphics[width=10cm]{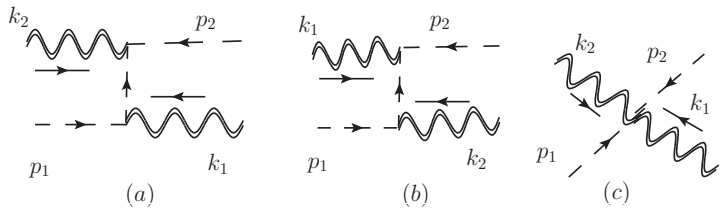}
  \caption{Compton scattering given by  the $t$-channel contribution in $(a)$, the $u$-channel contribution in $(b)$ and the contact term in $(c)$. }
  \label{fig:comptontree}
\end{figure}

Our starting point for deriving the gravitational Compton amplitude is the KLT expression from the previous section
\begin{equation}\label{e:Grav3}
i  M_s^{\rm tree} (p_1,p_2,k_1,k_2)={\kappa_{(4)}^2\over e^2}  {p_1\cdot
k_1\, p_1\cdot k_2\over k_1\cdot k_2}\, A_s^{\rm tree}(p_1,k_2,p_2,k_1)\, A_0^{\rm tree}(p_1,k_2,p_2,k_1)\,,
\end{equation}
where the gravity amplitude is expressed as a product of Yang-Mills amplitudes without a $s$-channel pole and we thus have no Yang-Mills diagrams involving the non-Abelian three-gluon vertex. This KLT representation of the gravitational Compton scattering
is key to the reduction of the amplitude to a product of QED amplitudes that we will consider in this section.
The color ordered-amplitude $A_s^{\rm tree}(p_1,k_2,p_2,k_1)$ represents the
scattering of a gauge boson off a spin $s$ matter field
depicted in figure~\ref{fig:comptontree}
\begin{equation}
  \label{e:Compton}
  A_s^{\rm tree}(p_1,k_2,p_2,k_1)  =e^2\left( {n_t^s\over p_1\cdot
      k_1}+{n_u^s\over p_1\cdot k_2}+ n_{ct}^s\right) \,,
\end{equation}
where $p_1^2=p_2^2=m^2$ are the momenta of the massive particles and
$k_1^2=k_2^2=0$ are the momentum of the gluons with all incoming momenta
$p_1+p_2+k_1+k_2=0$.

We will now explain that we can always express the amplitude
$A^{\rm tree}_s(p_1,k_2,p_2,k_1)$ solely in terms of
QED (abelian)  Compton scattering amplitudes.
The $t$- and $u$-channel
diagrams in figure~\ref{fig:comptontree}(a)-(b) are composed of
three-point amplitudes between two matter fields of the same spin~$s$
of the same flavor and one gauge boson.
The coupling of a matter field of spin~0 of the same species and one gauge boson is
given by
\begin{equation}
e\, (p_1-p_2)_\mu\, (2\pi)^4\, \delta(p_1+p_2+k_1)\,,
\end{equation}
or for particles of spins~$\frac12$ of the same species and one gauge boson
\begin{equation}
e\gamma_\mu\, (2\pi)^4\, \delta(p_1+p_2+k_1)\,.
\end{equation}
Finally the coupling  between two massive spin~1 fields of the
same species and one gauge boson is given by
\begin{equation}
-e\,(g_{\mu\nu} (k_1-p_2)_\rho+g_{\nu\rho}(p_2-p_1)_\mu +g_{\rho\mu}(p_1-k_1)_\nu) \, (2\pi)^4\, \delta(p_1+p_2+k_1)\,,
\end{equation}
where in all cases $e$ is the coupling constant.

There is no quartic coupling between two spinorial fields and one gauge boson and the
four-point interaction in fig.~\ref{fig:comptontree}(c)
between two scalars (without flavor changing)
and two gauge bosons
\begin{equation}
  -ie^2 g_{\mu\nu}   \, (2\pi)^4\,\delta(p_1+p_2+k_1+k_2)
\end{equation}
is the same in an non-Abelian as in an Abelian theory.

The four-point interaction between two massive vectors of the same
species and two gauge bosons is in an non-Abelian theory given by
\begin{multline}
-i e^2\,\sum_e [f_{abe}f_{e cc}
(g_{\mu\rho}g_{\nu\sigma}-g_{\mu\sigma}g_{\nu\rho})
+ f_{ace} f_{ebc} (g_{\mu\nu}g_{\rho\sigma}-g_{\mu\sigma}g_{\rho\nu})
+ f_{ace} f_{ebc} (g_{\mu\nu}g_{\sigma\rho} - g_{\mu\rho}g_{\sigma\nu})
] \cr\times\,(2\pi)^4\delta^4(p_1+p_2+k_1+k_2)  \,.
\end{multline}
By antisymmetry of the structure constant $f_{ecc}=0$  the
interaction reduces to
\begin{equation}
-i e^2\,(\sum_ef_{ace} f_{ebc})
[ (2g_{\mu\nu}g_{\rho\sigma}-g_{\mu\sigma}g_{\rho\nu} - g_{\mu\rho}g_{\sigma\nu})
] \times\,(2\pi)^4\delta^4(p_1+p_2+k_1+k_2)  \,,
\end{equation}
which has the same kinematic part as the Abelian one.

We can thus conclude that the amplitudes $A^{\rm tree}_s$ with
$s=0,\frac12$ appearing in the factorization of the gravity amplitudes
in~\eqref{e:Grav2} can be thought of as QED amplitudes for Compton scattering off
massive matter fields\footnote{The
representation of the massive gravitational Compton scattering of a
massive matter field of spin $s=0,1/2,1$ in terms of (Abelian) Compton amplitude was already noticed
in~\cite{Choi:1993xa,Choi:1993xa,Choi:1994ax}
and~\cite{Shim:1995ap}. It would be interesting to understand if this factorization using purely
Abelian interactions can be achieved with other types of gravitational amplitudes.
}.

The numerators of the QED  Compton amplitudes $A^{\rm tree}_s(p_1,k_2,p_2,k_2)$ are
given by
\begin{eqnarray}
\label{e:nt0}  n_t^0&=&2\epsilon_1\cdot p_1\,
  \,\epsilon_2 \cdot p_2\,, \\
\label{e:nt12}  n_t^{\frac12}&=&\frac12\, \bar u(-p_2)
  \slashed{\epsilon}_2(\slashed{p}_1+\slashed{k}_1+m)
  \slashed{\epsilon}_1 u(p_1)\,,\\
\label{e:nt1}  n_t^1&=&2[(h_1\cdot h_2)\,
(  \epsilon_1\cdot p_1)\, (\epsilon_2 \cdot  p_2) - h_1\cdot F_1\cdot F_2\cdot h_2 \cr
&-& (h_1 \cdot F_2 \cdot h_2)  \,
(  \epsilon_1\cdot p_1)
 -(h_1\cdot F_1\cdot h_2)  (\epsilon_2 \cdot p_2) ] \,,
\end{eqnarray}
and with similar expressions for $n_u^s$ with the exchange of $p_1$
and $p_2$ and finally\footnote{Notice that this is not a BCJ parameterization~\cite{BCJ}
because the numerators do not  satisfying a dual Jacobi identity. One
can define a set of BCJ numerators as $\tilde n_s^s=2(n_t^s+n_u^s)+t
n_{ct}^s$ and $\tilde n_t^s=-2n_t^s-t n_{ct}^s$ and $\tilde
n_u^s=-2n_u^s$, satisfying $\tilde n_s^s+\tilde n_t^s+\tilde n_u^s=0$.
Other expressions are possible, depending on the distribution of
contact terms amongst the pole terms.}
\begin{eqnarray}
\label{e:nct0}  n_{ct}^0&=&2\epsilon_1\cdot \epsilon_2\,, \\
\label{e:nct12}  n_{ct}^{\frac12}&=&0\,, \\
\label{e:nct1}  n_{ct}^1&=&-2h_1\cdot h_2 \, \epsilon_1\cdot \epsilon_2 \,.
\end{eqnarray}
We have here made use of the notation $h_1\cdot F_1\cdot
h_2=h_1^\mu  h_2^\nu F_{1\,\mu\nu}$ and $h_1\cdot F_1\cdot F_2\cdot
h_2=h_1^\mu F_{1\mu\rho} F_2{}^{\rho}{}_\nu h_2^\nu$ with
$F_{i\,\mu\nu}=k_{i\,\mu} \epsilon_{i\,\nu}-\epsilon_{i\,\nu} h_{i\,\mu}$ defining the
field-strengths of the photons.
With the given numerators factors we have checked that the spin~0
amplitude constructed from~\eqref{e:Grav2} correctly reproduces the covariant expression
in~\eqref{e:treeCov1}.

One important consequence of the factorization of the gravitational
Compton amplitude into a product of two Compton amplitudes is that it gives
a rationale for the value $g=2$  of the classical gyromagnetic momenta
for all types of matter fields, as shown
in ref.~\cite{Holstein:2006pq}.  An evaluation of the QED Compton amplitude
for massive particles shows that amplitude has a pole for $m=0$ with
residue $(g-2)^2$.
The two derivative nature of the
gravitational interaction forbids the presence of a singularity of the
gravitational Compton amplitude when the mass of the particles goes
to zero. Therefore the KLT relation
in~\eqref{e:Grav2} implies that the right hand side cannot have a pole in the
zero mass limit for generic values of the momenta. This implies the natural
classical value $g=2$ for all types of matter fields.

%-------------------------------------------------------------------------
\subsection{Helicity tree amplitudes for QED and Gravity}
\label{sec:helicity-trees}
%------------------------------------------------------------------------
\subsubsection{The QED amplitudes}
\label{sec:qed-amplitudes}
In this section we compare the QED amplitudes in~\eqref{e:Compton}
with the  scattering of two gluons off a  massive scalar derived using
the helicity formalism (see ref.~\cite{Badger:2005zh}). We use here $e^2=1$.  We have
\begin{equation}
  \label{e:4pointYM}
A_{0}^{\rm tree}(p_1,p_2,k_2^+,k_1^+)=-{m^2 \sq[k_1,k_2]^2\over
4(k_1\cdot k_2) \, (k_1\cdot p_1)},\qquad
A_{0}^{\rm tree} (p_1,p_2,k_2^-,k_1^+)={\spab[k_2,p_1,k_1]^2\over
  4(k_1\cdot k_2)\, (k_1\cdot p_1)}\,,
\end{equation}
with
\begin{eqnarray}
A_{0}^{\rm tree} (p_1,p_2,k_2^-,k_1^-)&=&(A_{0}^{\rm tree}
(p_1,p_2,k_2^+,k_1^+))^*\,,\cr
A_{0}^{\rm tree} (p_1,p_2,k_2^+,k_1^-)&=&(A_{0}^{\rm tree} (p_1,p_2,k_2^-,k_1^+))^*\,.
\end{eqnarray}
It is immediate to check that the Compton scalar amplitude $A^{\rm
  tree}_0(p_1,k_2,p_2,k_1)$ is related to the helicity amplitudes by the expected
monodromy relations $(p_1\cdot k_2)\,A^{\rm
  tree}_0(p_1,k_2,p_2,k_1)=(k_1\cdot k_2) A^{\rm tree}_0(p_1,p_2,k_2,k_1) $ and read
\begin{eqnarray}
  \label{e:Comptonhel}
  A_{0}^{\rm tree}(p_1,k_2^+,p_2,k_1^+)&=&-{m^2 \sq[k_1,k_2]^2\over
4 (p_1\cdot k_1) \, (p_1\cdot k_2)}\,,\cr
A_{0}^{\rm tree} (p_1,k_2^-,p_2,k_1^+)&=&{\spab[k_2,p_1,k_1]^2\over 4
\, (k_1\cdot p_1)(p_1\cdot k_2)}\,.
\end{eqnarray}
This expression is (although not manifestly) symmetric under the exchanges of $k_1$ and $k_2$ (with
flip of their helicities) and $p_1$ and $p_2$.

%-----------------------------------------------------------------------
\subsubsection{The gravity amplitudes}
\label{sec:gravity-amplitudes}
Using the relation in~\eqref{e:Grav1} we can write the expression
for the four-point
amplitudes for the emission of two gravitons. In this situation, we have
\begin{eqnarray}
  \label{e:4pointGrav}
iM_{0}^{\rm tree}(p_1,p_2,k_1^+,k_2^+) &=&{\kappa_{(4)}^2\over 16}\,
{1\over (k_1\cdot k_2) }\,
{m^4 \sq[k_1,k_2]^4\over (k_1\cdot p_1)(k_1\cdot p_2)} \,, \cr
iM_{0}^{\rm tree}(p_1,p_2,k_1^-,k_2^+)&=&{\kappa_{(4)}^2\over16}\,{1\over (k_1\cdot k_2)}\,
{ \spab[k_1,p_1,k_2]^2\spab[k_1,p_2,k_2]^2\over (k_1\cdot p_1)(k_1\cdot p_2)}\,,
\end{eqnarray}
with 
\begin{equation} iM_{0} ^{\rm tree}(p_1,p_2,k_1^-,k_2^-)=(iM_{0} ^{\rm
tree}(p_1,p_2,k_1^+,k_2^+))^*\,,\end{equation}
 and 
\begin{equation}iM_{0} ^{\rm tree}(p_1,p_2,k_1^+,k_2^-)=(iM_{0} ^{\rm
tree}(p_1,p_2,k_1^-,k_2^+))^*.\end{equation}
We have checked that these expressions match the covariant ones and the
expression obtained from~\eqref{e:Grav2}.  The
massive
amplitude $M_{0}^{\rm tree}(p_1,p_2,k_1^+,k_2^-)$ reproduces the one
given in~\cite[eq.~(5.4)]{Dunbar:1995ed} and its massless limit
reproduces the results
of~\cite[eq.~(4.5)]{Dunbar:1995ed}. We note that using the KLT factorization
property to construct the amplitudes that go into the cut avoids having to
deal with tensor contractions of the complicated triple graviton vertex, which is
a normal tedious feature of any off-shell Feynman diagram computation.

%%%%%%%%%%%%%%%%%%%%%%%%%%%%%%%%%%%%%%%%%%%%%%%%%%%%%%%%%%%%%%%%%
\section{The one-loop amplitude in the helicity formalism}
\label{sec:helicity}
\begin{figure}[ht]
  \centering\includegraphics[width=6cm]{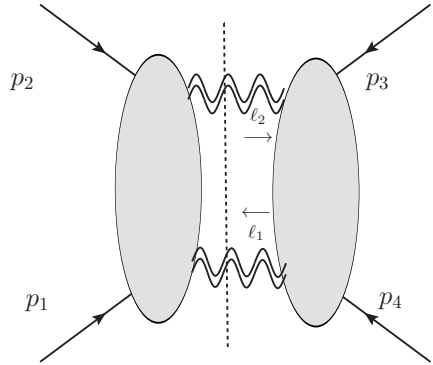}
  \caption{ The cut  considered. The loop momenta
    are flowing clockwise. And the on-shell conditions are
    $\ell_1^2=0$ and $\ell_2^2=(\ell_1+p_1+p_2)^2=0$.
Solid lines are massive and wiggly lines are massless.
\label{fig:cuts} }
  \end{figure}
In this section, we obtain the non-analytic terms that give the leading
classical and quantum corrections to the scattering potential for QED and for
general relativity. For this purpose we do not need to reconstruct the
full amplitude, but only identify those terms in the cut that lead to non-analytic
contributions, {\it i.e.} $C^{NP}$, the classical
correction from general relativity, and $C^{GQ}$, the quantum gravity correction to
Newton's potential in~\eqref{e:Vq}. We obtain these respectively from the coefficients
of the non-analytic $1/\sqrt{-q^2}$ and $\log(-q^2)$ contributions in cut.

To extract the non-analytic parts of the amplitude, we will proceed as in ref.~\cite{Bern:1994cg}. Instead of
evaluating the phase-space integrals directly we simply reinstate the off-shell cut
propagators but impose strictly the on-shell cut condition everywhere in the numerator.
We thus evaluate the following types of expressions
  \begin{equation}
  \label{e:dischel}
\left.  iM^{\rm 1-loop}\right|_{disc}=\int {d^D\ell\over (2\pi)^D}\,
 {\sum_{\lambda_1,\lambda_2} M^{\rm
    tree}_{\lambda_1\lambda_2}(p_1,p_2,-\ell_2^{\lambda_2},\ell_1^{\lambda_1})
 (M^{\rm
    tree}_{\lambda_1\lambda_2}(p_3,p_4,\ell_2^{\lambda_2},-\ell_1^{\lambda_1}))^* \over \ell_1^2 \ell_2^2}\Big|_{cut}
  \,,
\end{equation}
with $\ell_1^2=\ell_2^2=0$ and where $\lambda_1$ and $\lambda_2$ are the
helicities of the massless particles (gravitons/photons) across the  cut. In this formula, we are
using the notation $|_{cut}$ to indicate the cut is taken in this
integral. Whenever we discuss the discontinuity singularity it is understood
that we are on the cut, although we will not explicitly indicate this
in the integral for simplicity. This procedure allows us to directly identify the box,
triangle and bubble integral functions which contribute
to the amplitude, and use them to identify the non-analytic terms which we are seeking\footnote{
By considering only this two-particle discontinuity across the massless momenta,
we do not have enough
information to reconstruct the full amplitude. To achieve this, we would
need to consider all the discontinuities across the massive legs and evaluate
the cut to all orders in $\epsilon$ with $D\equiv4-2\epsilon$.
However, the discontinuities
across the massive propagators will not contribute to the leading order massless
threshold, nor will higher order terms from an $\epsilon$ expansion of the cut.
Thus we will ignore all these contributions here as they are not important for our analysis.}.

We will illustrate our discontinuity cut method by first calculating the case of the Coulomb
potential. Here the cut is a little simpler and it is easier to demonstrate the techniques.
In the next subsection we will then use the cut technique in the case of pure gravity.

%-------------------------------------------------------------------------
\subsection{The one-loop correction to Coulomb potential}
\label{sec:oneloopQED}
In this section we will compute the quantum correction to the Coulomb
potential between two spin~0 particles of the same charge but non-zero masses  $m_1$
and $m_2$.

We are constructing the one-loop amplitude by computing its
discontinuity
cut across the massless  photon lines (double wavy-line in figure~\ref{fig:cuts}).
We not are interested in reconstructing the full one-loop amplitude but only
the parts that contain the  infra-red logarithms and square-root contributions.

In the cut in Eq.~\eqref{e:dischel} we have the following on-shell  kinematic relations
 $p_1+p_2+p_3+p_4=0$,  $p_1^2=p_2^2=m_1^2$ and
$p_3^2=p_4^2=m_2^2$. We define the momentum transfer $q$ from $q=p_1+p_2=-(p_3+p_4)$.
We have in the static non-relativistic limit $p_1, -p_2\simeq (m_1,\vec 0)$ and
$p_4,-p_3\simeq (m_2,\vec 0)$,  and furthermore that (in the mostly minus metric)
\begin{eqnarray}\label{e:NRlimit}
s&=&(p_1+p_2)^2\simeq -{\vec q}^{\;2}\,,\cr
t&=&(p_1+p_4)^2\simeq (m_1+m_2)^2\,,\\
\nn  u&=&(p_1+p_3)^2\simeq (m_1-m_2)^2+{\vec q}^{\;2}\,.
\end{eqnarray}
The tree-level helicity amplitudes are given in~\eqref{e:Comptonhel} hence
 the discontinuity of the one-loop amplitude takes the form
\begin{equation}
  \label{e:singcut}
\left.  iM^{\rm 1-loop}\right|_{disc}={e^4\over 16}\int {d^D\ell\over (2\pi)^D}\,
 {  \cN\over \ell_1^2 \ell_2^2\, \prod_{i=1}^4 (p_i\cdot \ell_1)}
  \,.
\end{equation}
We deduce that
\begin{eqnarray}\label{e:scutreduc}
  {1\over \ell_1\cdot p_1\,\ell_1\cdot p_2}&=&-{2\over s}\, \left({1\over \ell_1\cdot
    p_1}+{1\over \ell_1\cdot p_2}\right)\,,  \cr
 {1\over \ell_1\cdot p_3\,\ell_1\cdot p_4}&=&{2\over s}\, \left({1\over \ell_1\cdot
    p_3}+{1\over \ell_1\cdot p_4}\right) \,,
\end{eqnarray}
using that $q=p_1+p_2=\ell_2-\ell_1=-p_3-p_4$ and $\ell_1\cdot q=-s/2$.
This allows us to express the one-loop cut as a sum of integrals with
numerator $\cN$
\begin{equation}
  \label{e:singcut2QED}
\left.  iM^{\rm 1-loop}\right|_{disc}=-{e^4\over 4}\sum_{i=1}^2\sum_{j=3}^4\int {d^D\ell\over (2\pi)^D}\,{\cN\over s^2 \ell_1^2 \ell_2^2
   (p_i\cdot \ell_1) (p_j\cdot \ell_1)}
  \,.
\end{equation}
where we will distinguish between the cases of the photons having the same helicity on each side of the cut (this is traditionally in the literature called a singlet contribution) or opposite helicity (called a non-singlet contribution).

For the singlet cut  the numerator is given by
\begin{equation}
  \label{e:singletQED}
\cN^{\rm singlet}= m_1^2 m_2^2 s^2\,.
\end{equation}
Giving a contribution from the singlet cut of only scalar boxes
\begin{equation}\label{e:singletAmpQED}
  M^{\rm singlet}  =-e^4\, 2m_1^2m_2^2 (I_4(s,t)+I_4(s,u))\,,
\end{equation}
Here we have in $D=4-2\epsilon$ using the normalization of ref.~\cite{BjerrumBohr:2002kt} that
\begin{eqnarray}
  \label{e:scalarboxes}
  I_4(s,t)&=&\frac{1}{i}\int {d^D\ell\over(2\pi)^D} \, {1\over \ell^2 (\ell+q)^2\
    ((\ell+p_1)^2-m_1^2) ((\ell-p_4)^2-m_2^2)}\,, \cr
I_4(s,u)&=&\frac{1}{i}\int {d^D\ell\over(2\pi)^D} \, {1\over \ell^2 (\ell+q)^2
    ((\ell+p_1)^2-m_1^2) ((\ell-p_3)^2-m_2^2)}\,,
\end{eqnarray}
In the non-relativistic limit and to leading order in $q^2$, we have the logarithmic terms~\cite{BjerrumBohr:2002kt} 
\begin{eqnarray}\label{e:boxNRlimit}
\nn I_4(s,t)+I_4(s,u)&=&{\log({\vec q}^2)\over 96\pi^2m_1^2m_2^2}\\
  I_4(s,t)-I_4(s,u)&=&-{\log({\vec q}^2)\over 8\pi^2m_1m_2\, {\vec q}^2}\,.
\end{eqnarray}
Note that the box diagram also contains an imaginary part which survives in the non-relativistic limit
\begin{equation}
\Im\textrm{m}~I_4(s,t) = - \frac{1}{16\pi(m_1+m_2)q^2 p}
\end{equation}
where 
\begin{equation}
p = \left[\frac{(s-(m_1+m_2)^2))(s-(m_1-m_2)^2}{4s}\right]^{1/2}
\end{equation}
Weinberg \cite{Weinberg:1965nx} has shown that this leads to an overall
phase on the scattering amplitude and does not contribute to observables. For this reason we do not include it here, nor in subsequent discussions. 
With these results, the singlet cut amplitude in~\eqref{e:singletAmpQED} in
the non-relativistic limit gives
\begin{equation}\label{e:SingletQED}
  M^{\rm singlet}(q)\simeq -{e^4\over(4\pi)^2}\, {1\over3} \log({\vec q}^{\;2}) \,,
\end{equation}
to leading order in $q^2\sim - \vec
q\cdot \vec q$.

For the non-singlet cut contribution the numerator is given by
\begin{equation}
  \label{e:nonsingletQED}
\cN^{\rm non-singlet}=  \frac12( \tr_-(\ell_2 p_1\ell_1p_3)^2 + \tr_+(\ell_2 p_1\ell_1p_3)^2) \,,
\end{equation}
where the traces are defined by
$
  \tr_\pm(abcd)=2 (a\cdot b\, c\cdot d - a\cdot c \, b\cdot d+a\cdot d\,
  b\cdot c)\pm 2i \epsilon^{\mu\nu\rho\sigma} a_\mu b_\nu c_\rho d_\sigma\,.
$
Expanding the traces we see that one can rewrite the numerator in terms of two contributions
$\cN^{\rm non-singlet}\equiv\mathcal E^2-4\mathcal O$ where
\begin{eqnarray}
\mathcal  E&:=& 2 (\ell_1\cdot p_1 \, \ell_2\cdot p_3-\ell_1\cdot \ell_2 \,
   p_1\cdot p_3+\ell_1\cdot p_3\, \ell_2\cdot p_1)\,,\cr
  \mathcal O&:=&(\epsilon^{\mu\nu\rho\sigma} \ell_{1\,\mu}
  p_{1\,\nu}\ell_{2\,\rho}p_{3\,\sigma})^2 \,.
\end{eqnarray}
This leads in the non-relativistic approximation to a rather simple
form for the numerator
\begin{equation}
  \label{e:NumeratorQEDNR}
 \mathcal N^{\rm non-singlet}\simeq (s\, m_1^2 + 4 (p_1\cdot
 \ell_1)^2)\, (s\, m_2^2  + 4 (p_4\cdot \ell_1)^2)\,.
\end{equation}

Evaluating the contributions from the non-singlet cut (in the non-relativistic limit) lead
to the following combinations of scalar box, triangle and bubble integrals to leading order
\begin{multline}\label{e:nonsingletAmpQED}
  M^{\rm non-singlet}=- e^4\, \Big( 2 m_1^2m_2^2 (I_4(s,t)+I_4(s,u))+
    m_1^2 (I_3(p_1,q,m_1)+I_3(p_2,q,m_2))\cr+ m_2^2
    (I_3(-p_3,q,m_2)+I_3(-p_4,q,m_2))+ I_2(q)\Big)\,.
\end{multline}
The scalar triangle and bubbles integrals are defined following the
conventions of ref.~\cite{BjerrumBohr:2002kt}
\begin{eqnarray}
  \label{e:scalartribub}
  I_3(p,q,m)&:=&{1\over i} \int {d^D\ell\over (2\pi)^D} \, {1\over
    \ell^2(\ell+q)^2 ((\ell+p)^2-m^2)}\,,\cr
I_2(q)&:=& {1\over i}\int {d^D\ell\over (2\pi)^D} \, {1\over
    \ell^2(\ell+q)^2}\,.
\end{eqnarray}
Where we have, in the non-relativistic limit,
\begin{eqnarray}
  \label{e:scalartribubNR}
  I_3(p,q,m)&\simeq&-{1\over 32\pi^2m^2}\,\left(\log(-q^2)+S(m)\right)\,, \\
I_2(q)&\simeq& {1\over 16\pi^2}\,\log(-q^2) \,,
\end{eqnarray}
defining $S(m)= -\pi^2 m/|{\vec q}|$.

Thus the contribution from the non-singlet cut amplitude in~\eqref{e:nonsingletAmpQED} yields
in the non-relativistic limit $w\to0$ (to the first order in $q^2\simeq - {\vec q}^{\;2}$)
\begin{equation}\label{e:NonSingletQED}
  M^{\rm non-singlet}(q) \simeq {e^4\over(4\pi)^2}\,\left({8\over3} \,\log(-q^2)
    -\pi^2{m_1+m_2\over |\vec q|} \right)
\end{equation}
Summing~\eqref{e:SingletQED} and~\eqref{e:NonSingletQED}  we obtain the total amplitude
\begin{equation}
  M^{\rm non-rel.}(q)\simeq {e^4\over(4\pi)^2}\,\left({7\over3} \,\log({\vec
    q}^{\;2}) -  \pi^2 {m_1+m_2\over |\vec q|}\right)\,,
  \end{equation}
and the one-loop correction to the non-relativistic potential is given
by
\begin{equation}
  V^{\rm one-loop}(q)=  {M^{\rm non-rel.}(q)\over 4m_1m_2}=  {e^4\over8\pi^2m_1m_2}\,\left( {7\over3} \,\log({\vec
    q}^{\;2})  - \pi^2 {m_1+m_2\over |\vec q|}\right)\,.
\end{equation}
 This reproduces the result of~\cite[eqs.~(4.50a), (4.51a),
 (4.54)]{Feinberg:1988yw} and~\cite{Holstein:2008sw}, although we want
 to point out the huge simplicity of our cut derivation.

%-------------------------------------------------------------------------
\subsection{The one-loop correction to Newton potential}
\label{sec:oneloophel}
In this section we will perform the evaluation of the correction to the
Newton potential using the on-shell cut in the helicity formalism. This computation will as expected not require any ghost
contributions.

Proceeding as in the QED case, the cut discontinuity of the  amplitude can be
expressed as  a sum of  integrals with numerator $\cN$
\begin{equation}
  \label{e:singcut2grav}
\left. i M^{\rm 1-loop}\right|_{disc}=-{\kappa_{(4)}^4\over 16\,s^4}\sum_{i=1}^2\sum_{j=3}^4\int {d^D\ell\over (2\pi)^D}\,{\cN\over \ell_1^2 \ell_2^2
   (p_i\cdot \ell_1) (p_j\cdot \ell_1)}
  \,.
\end{equation}
We will  evaluate the amplitude  in the static non-relativistic limit~\eqref{e:NRlimit}.

As in the QED case, we will here as well distinguish between the cases of the graviton having the same helicity on each side of the cut (singlet) or opposite helicity (non-singlet), and we separate the numerator factor $\cN$ in these two contributions.

The singlet-cut numerator is easily evaluated and gives
\begin{equation}
  \label{e:Nsinglet}
\mathcal N^{\rm singlet}=2 {m_1^4\,m_2^4\, (\ell_1\cdot \ell_2)^4}=
  {m_1^4m_2^4\over 8}\, s^4\,,
\end{equation}
therefore its contribution to the one-loop amplitude is given by
scalar boxes only
\begin{equation}
  M^{\rm singlet}(q)=-{\kappa_{(4)}^4\over16}\, m_1^4m_2^4 \, (I_4(s,t)+I_4(s,u))\,.
\end{equation}
This is readily evaluated in the non-relativistic limit to give
\begin{equation}\label{e:MBoxesSinglet}
  M^{\rm singlet}(q)\simeq -G_N^2\, {4m_1^2m_2^2\over 3}\,\log({\vec q}^{\;2})\,,
\end{equation}
where we have made use of the relation $\kappa_{(4)}^2=32\pi G_N$.

The numerator for the non-singlet cut contribution is evaluated to
\begin{equation}
  \label{e:gg-nonsinglet}
  \cN^{\rm non-singlet}=\frac12\left((\tr_-(\ell_1p_1\ell_2p_3))^4+ (\tr_+(\ell_1p_3\ell_2p_1))^4\right)\,.
\end{equation}
The evaluation of this contribution is a bit more involved since the
expression contains integrals with up to eight powers of loop momentum in
the numerator. We note that in the gravity case the cut is not the square of the QED cut but
the sum of the squares of the corresponding QED terms in the cut.

Decomposing the trace as in the QED case (keeping only terms that give a contribution
in the non-relativistic limit) the numerator factor takes
the form
\begin{eqnarray}\label{e:expandedgg}
  \mathcal N^{\rm non-singlet}&=& ((\mathcal E^2-4\mathcal
  O)^2-16\mathcal E^2\mathcal O)\cr
&=&16 s^2 (m_2\, p_1\cdot \ell_1 - m_1 \,p_4\cdot \ell_1)^4 \cr
&+ &
 24 s (m_2\, p_1\cdot \ell_1 - m_1\, p_4\cdot \ell_1)^2 (m_1 m_2 s +
    4 p_1\cdot \ell_1 p_4\cdot \ell_1)^2\cr
& +& (m_1 m_2 s + 4 p_1\cdot \ell_1 p_4\cdot \ell_1)^4\,.
\end{eqnarray}
In the non-relativistic limit evaluating the discontinuity cut integrals leaves us with a sum of scalar
boxes, scalar, linear and quadratic triangles and bubbles integral functions ranging from
scalar to quartic, {\it i.e.}
\begin{equation}
  M^{\rm non-singlet}(q)= M^{\rm non-singlet}_{\rm boxes} (q)+   M^{\rm non-singlet}_{\rm triangles}(q) +   M^{\rm non-singlet}_{\rm bubbles}(q)\,.
\end{equation}
To the leading order in the non-relativistic limit, we have scalar box integral functions $M^{\rm boxes}$ given by
\begin{equation}
  \label{e:MboxesNonSinglet}
   M^{\rm non-singlet}_{\rm boxes} (q)=-{\kappa_{(4)}^4\over8}\,\left(
     m_1^4m_2^4    (I_4(s,t)+I_4(s,u))+2m_1^3 m_2^3\, s\,(I_4(s,t)-I_4(s,u)) \right)\,.
\end{equation}
This gives in the non-relativistic limit (using~\eqref{e:boxNRlimit})
\begin{equation}
  \label{e:MboxesNonSingletNR}
  M^{\rm non-singlet}_{\rm boxes}\simeq -G_N^2\, {100 m_1^2m_2^2\over
    3}\,\log({\vec q}^{\;2})\,.
\end{equation}
where again we have discarded the imaginary part of the box diagram. 
For the triangles, we have integrals from scalars up to quadratic terms $M^{\rm triangles}$,
\begin{eqnarray}
&& M^{\rm non-singlet}_{\rm triangles}=-{\kappa_{(4)}^2\over16}\,\Big[\\
&&6 m_1^4 m_2^2(   I_3(p_1)+I_3(p_2)) +6   m_1^2 m_2^4 (I_3(-p_3)+   I_3(-p_4))\cr
&-&2 m_1^4 (   I_3(p_1,\{p_4\})+ I_3(p_2,\{p_4\}))+8 m_1^3   m_2 (I_3(p_1,\{p_4\})- I_3(p_2,\{p_4\}))\cr
&+&2 m_2^4   (I_3(-p_3,\{p_1\})+ I_3(-p_4,\{p_1\}))+8 m_1 m_2^3   (I_3(-p_3,\{p_1\})-I_3(-p_4,\{p_1\}))\cr
&+&4 m_1^2 (I_3(p_1,\{p_4,p_4\})+   I_3(p_2,\{p_4,p_4\}))
+4 m_2^2(   I_3(-p_3,\{p_1,p_1\})+   I_3(-p_4,\{p_1,p_1\}))\cr
\nn&+&{4\over q^2}\, \Big(m_1^4 (I_3(p_1,\{p_4,p_4\})+   I_3(p_2,\{p_4,p_4\}))+m_2^4
   (I_3(-p_3,\{p_1,p_1\})+   I_3(-p_4,\{p_1,p_1\}))\Big)\Big]\,,
\end{eqnarray}
with  linear and quadratic triangles defined via
\begin{equation}
I_3(p,q,m;\{K_1,\dots,K_r\}):={1\over i} \int {d^D\ell\over(2\pi)^D} \, {
 \prod_{i=1}^r \ell\cdot K_i\over \ell^2(\ell+q)^2
(  (\ell+p)^2-m^2)}\,,
\end{equation}
where we have $r=0$ for scalar triangles, $r=1$ for linear triangles
and $r=2$ for quadratic triangles.  We use here the short hand
notation that $I_3(p_r,\cdots )=
I_3(p_r,q,m_1,\cdots)$ for $r=1,2$ and $I_3(-p_r,\cdots )=
I_3(-p_r,q,m_2,\cdots)$ for $r=3,4$.

Taking the non-relativistic limit leaves us with
\begin{eqnarray}
  \label{e:hightribub}
  I_3(p,q,m; K_1)&\simeq&{1\over 32\pi^2m^2}\,\left[(K_1\cdot
    p)\left(-1-{q^2\over2m^2}\right)\log(-q^2)
+K_1\cdot q\, (\log(-q^2)+\frac12\, S(m))\right] \,,\cr
I_3(p,q,m; K_1,K_2)&\simeq&{1\over 32\pi^2m^2}\,\left[(K_1\cdot
  q)(K_2\cdot q) (-\log(-q^2)-\frac38 S(m))\right.\cr
&-&(K_1\cdot p)(K_2\cdot p) \, {q^2\over8m^2}
(4\log(-q^2)+S(m))\cr
&+&((K_1\cdot q)(K_2\cdot p)+(K_1\cdot p)(K_2\cdot
q))\,\left({q^2+m^2\over2m^2} \log(-q^2)+{3q^2\over16m^2}S(m)\right)\cr
&+&\left. \frac18  K_1\cdot K^2\, q^2\, (2\log(-q^2)+S(m))\right]\,,
\end{eqnarray}
so that
\begin{equation}
  \label{e:MtrianglesNonSinglet}
   M^{\rm non-singlet}_{\rm triangles}(q) \simeq G_N^2\, m_1^2m_2^2\,\left(
   120\,\log(\vec q\cdot \vec
   q)-24\pi^2\, {m_1+m_2\over|\vec q|}\right)\,.
\end{equation}

To the leading order in the non-relativistic limit, the bubble
contribution $M^{\rm non-singlet}_{\rm bubble}$ is given by
\begin{eqnarray}
 M^{\rm non-singlet}_{\rm
   bubbles}&=&-{\kappa_{(4)}^4\over16}\,\Big[{16\over s^2}
 I_2(q,\{p_1,p_1,p_4,p_4\})\\
&-&4 \Big(3 m_1^2 m_2^2 I_2(q) -m_2  (2 m_1+3 m_2) I_2(q,\{p_1\})+ m_1 (3m_1+2m_2 )
   I_2(q,\{p_4\})\cr
&+& I_2(q,\{p_1,p_1\})+   I_2(q,\{p_4,p_4\})+3  I_2(q,\{p_1,p_4\})  \Big)\cr
&+&{8\over s}\, \Big(3(m_2^2  I_2(q,\{p_1,p_1\}) +m_1^2  I_2(q,\{p_4,p_4\}))-4 m_1 m_2
   I_2(q,\{p_1,p_4\})\cr
&+& I_2(q,\{p_1,p_4,p_4\})- I_2(q,\{p_1,p_1,p_4\})\Big)\,,
\end{eqnarray}
where
\begin{equation}
  I_2(q,\{K_1,\cdots,K_r\}):={1\over i}\int {d^D\ell\over(2\pi)^D} \,
{\prod_{i=1}^r \ell\cdot K_r\over \ell^2 (\ell+q)^2}\,,
\end{equation}
with $r=0,1,2,3,4$. The bubble integrals are all given by
\begin{equation}
    I_2(q,\{K_1,\cdots,K_r\}) = I_2(q) \, P_r(  q^2) + \textrm{rational~part}\,,
\end{equation}
where $I_2(q)$ is the scalar bubble function given
in~\eqref{e:scalartribub} and $P_r(q,K_1,\dots,K_r)$ is a polynomial. The rational
part does not contribute to our analysis.
The polynomials are given by
\begin{eqnarray}
  P_1(q,K_1) &=&-{q \cdot K_1\over2}\,,\\[.5ex]
\nn P_2(q,K_1,K_2)&=&{1\over 12}\,
(4q\cdot K_1 \, q\cdot K_2   - q^2K_1\cdot K_2)\,,\\[.5ex]
\nn P_3(q,K_1,K_2,K_3)&=&\frac{1}{24} \Big(q^2 \,(K_1 \cdot K_3
   K_2 \cdot q+K_1 \cdot K_2 K_3 \cdot q+K_1 \cdot q
   K_2 \cdot K_3)\\
\nn&-&6 K_1 \cdot q K_2 \cdot q K_3 \cdot q)\Big)\,,\\[.5ex]
\nn P_4(q,K_1,K_2,K_3,K_4)&=&\frac{1}{240}
\Big((q^2)^2 (K_1 \cdot K_4 K_2 \cdot K_3+K_1 \cdot K_3
   K_2 \cdot K_4+K_1 \cdot K_2 K_3 \cdot K_4)\\
&-&6q^2 (
   K_1 \cdot q K_2 \cdot q K_3 \cdot K_4+ K_1 \cdot K_4 K_2 \cdot q
   K_3 \cdot q+ K_1 \cdot K_3 K_2 \cdot q K_4 \cdot q+\cr
&&   K_1 \cdot K_2 K_3 \cdot q K_4 \cdot q+ K_1 \cdot q
   K_2 \cdot K_4 K_3 \cdot q+K_1 \cdot q  K_2 \cdot K_3 K_4 \cdot
   q)\cr
\nn &+&48
   K_1 \cdot q K_2 \cdot q K_3 \cdot q K_4 \cdot q\Big)\,.
\end{eqnarray}
Leaving us with
\begin{equation}
  \label{e:MbubblesNonSinglet}
   M^{\rm non-singlet}_{\rm bubbles} (q)=G_N^2\, {788m_1^2m_2^2\over
     15}\,\log({\vec q}^{\;2})\,.
\end{equation}
Thus the total contribution is given by
summing~\eqref{e:MBoxesSinglet}, \eqref{e:MboxesNonSingletNR},~\eqref{e:MtrianglesNonSinglet}
and~\eqref{e:MbubblesNonSinglet} yielding
\begin{equation}
  \label{e:totalhelicity}
  M^{\rm total}(q)=G_N^2 \,4m_1^2m_2^2\left(-6\pi^2 {m_1+m_2\over|\vec q|}+{41\over 5}\,\log({\vec q}^{\;2})\right)\,,
\end{equation}
leading to the one-loop correction to the non-relativistic  potential
\begin{equation}
    V^{\rm one-loop}(q)=  {M^{\rm total}(q)\over 4m_1m_2}= G_N^2 m_1m_2\,\left(-6\pi^2 {m_1+m_2\over|\vec q|}+{41\over 5}\,\log({\vec q}^{\;2})\right)\,.
\end{equation}
This matches refs.~\cite{BjerrumBohr:2002kt,
  Khriplovich:2002bt}. We point out that other
 computations can be carried out with much greater ease using the cut method as well,
 for example the mixed electromagnetic-gravitational scattering case,
 previously computed in refs.~\cite{BjerrumBohr:2002sx,Faller:2007sy}.

%-------------------------------------------------------------------------
\section{The  one-loop amplitude in harmonic gauge}
\label{sec:harmonic}
We can also use the discontinuity cut technique to evaluate the potential
using the covariant notation, in harmonic gauge. This
has two interesting features. One is that this gauge requires ghost fields, and we
will see that the discontinuity from the ghosts must be added in order to obtain the full result.
In addition, this calculation lets us make direct contact with the Feynman diagram approach in
harmonic gauge~\cite{BjerrumBohr:2002kt, Khriplovich:2002bt}. We will describe in this section
how one can compare with the individual diagrams of the effective field theory
calculation.
%
%------------------------------------------------------------------------
\subsection{The graviton and ghost contributions}
Our starting point is the tree-level amplitude which takes the generic form
\begin{equation}
  \label{e:MtreeGen}
  M^{\rm tree}(p_1,p_2,k_1,k_2)= M^{\rm
    tree}_{\mu\nu,\rho\sigma}(p_1,p_2,k_1,k_2) \epsilon^{\mu\nu}(k_1) \epsilon^{\rho\sigma}(k_2)\,.
\end{equation}
When we take the discontinuity across the massless graviton lines we use the
harmonic gauge polarization sum $\cP_{\alpha\beta,\gamma\delta}$ given in Eq. \ref{e:defP}.
This yields the expression for the on-shell discontinuity (in $D=4-2\epsilon$ dimensions)
\begin{equation}
  \label{e:oneloopcut}
\left.  iM^{1-loop}\right|_{disc}=\int {d^D\ell\over (2\pi)^D}\,
 {M^{\rm
    tree}_{\mu\nu,\rho\sigma}(p_1,p_2,-\ell_2,\ell_1)
  \mathcal \cP^{\mu\nu,\alpha\beta} \mathcal \cP^{\rho\sigma,\gamma\delta} (M^{\rm
    tree}_{\alpha\beta,\gamma\delta}(p_4,p_3,\ell_2,-\ell_1))^* \over \ell_1^2 \ell_2^2}
  \,.
\end{equation}
A significant simplification in evaluating the discontinuity across the cut
in~\eqref{e:oneloopcut} is due to the following remarkable identities
noticed in~\cite{BjerrumBohr:2002kt}
\begin{eqnarray}
  \tau_{2\,\mu\nu,\rho\sigma}(p_1,p_2,m)
 \cP^{\mu\nu}{}_{\alpha\beta}
  \cP^{\rho\sigma}{}_{\gamma\delta}&=&
  \tau_{2\,\alpha\beta,\gamma\delta}(p_1,p_2,m) \,,  \\
\tau^{\rho\sigma}_{3\,\mu\nu,\rho\sigma}(k_1,k_2,q)
  \cP^{\mu\nu}{}_{\alpha\beta}
  \cP^{\rho\sigma}{}_{\gamma\delta}&=&
  \tau^{\rho\sigma}_{3\,\alpha\beta,\gamma\delta}(k_1,k_2,q)  \,.
\end{eqnarray}
The identification of the boxes, triangles and bubbles is not as neat
as in the helicity approach, and we do not display the intermediate
formulas. Performing the index contraction with Mathematica and taking the
non-relativistic limit as described in~\cite{BjerrumBohr:2002kt} we
obtain for the contribution of the cut in eq.~\eqref{e:oneloopcut}
\begin{equation}
  \label{e:oneloopNR}
iM^{\rm disc}(q)\simeq G_N^2\,4 m_1^2m_2^2\,\left(-{26\over3} \log({\vec q}^{\;2}) -6 \pi^2{
m_1+m_2\over |\vec q|}\right)\,.
\end{equation}
\begin{figure}[ht]
  \centering
  \includegraphics[width=5cm]{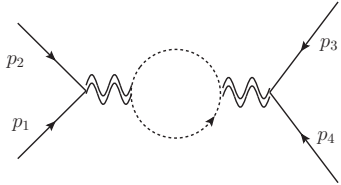}
  \caption{The ghost contribution from the vacuum polarization of the graviton}
  \label{fig:ghost}
\end{figure}
Since we used the harmonic gauge in this covariant computation we
need to include the extra graph of figure~\ref{fig:ghost}  from the contribution of the ghost to
the vacuum polarization of the graviton.
The ghost Lagrangian for the de Donder harmonic gauge used in this
work reads~\cite{'tHooft:1974bx,Veltman:1975vx}
\begin{equation}
  \label{eq:14}
  \mathcal S^{\rm  ghost}= \int d^4x \, \sqrt{g}\, \eta^{*\,\mu}\,
  (\nabla^\lambda \nabla_\lambda \eta_\mu+ \nabla^\lambda \nabla_\mu
  \eta_\lambda-\nabla_\mu \nabla_\lambda\eta^\lambda)\,.
\end{equation}
Evaluating the graph in figure~\ref{fig:ghost} leads to the
contribution in the non-relativistic limit
\begin{equation}\label{e:ghost}
  M^{\rm ghost}(q)\simeq G_N^2\,  {1012\over15} m_1^2m_2^2 \log({\vec q}^{\;2})\,.
\end{equation}
Summing the contributions in~\eqref{e:oneloopNR} and~\eqref{e:ghost}
leads to the result given by the helicity computation~\eqref{e:totalhelicity} and verifies again ref.~\cite[eq.~(44)]{BjerrumBohr:2002kt}
\begin{equation}
iM^{\rm 1-loop}( q)  \simeq  G_N \,4m_1^2m_2^2\left( -6 \pi^2{m_1+m_2\over|\vec
    q|}
    +{41\over 5}\,\log({\vec q}^{\;2})\right)\,.
\end{equation}
By way of comparison, we note that the helicity amplitude calculation of the previous
section corresponds to a sum over the physical helicities
\begin{equation}
  \label{e:oneloopcutS}
\left.  iM^{1-loop}\right|_{disc}=\int {d^D\ell\over (2\pi)^D}\,
 {M^{\rm
    tree}_{\mu\nu,\rho\sigma}(p_1,p_2,-\ell_2,\ell_1)
  \mathcal S^{\mu\nu,\alpha\beta} \mathcal S^{\rho\sigma,\gamma\delta} (M^{\rm
    tree}_{\alpha\beta,\gamma\delta}(p_4,p_3,\ell_2,-\ell_1))^* \over \ell_1^2 \ell_2^2}
  \,,
\end{equation}
where $\mathcal S_{\mu\nu,\rho\sigma}$ arises from the axial-gauge polarization sum
\begin{equation}\label{e:defSS}
\mathcal S_{\mu\nu,\rho\sigma}:=\sum_{\lambda=\pm1} \epsilon^{\lambda\lambda}_{\mu\nu}(k) (
  \epsilon^{\lambda\lambda}_{\rho\sigma}(k) )^*
=\frac12\,\left( S_{\mu\rho}S_{\nu\sigma}+S_{\nu\rho}S_{\mu\sigma}-S_{\mu\nu}S_{\rho\sigma}\right)\,,
\end{equation}
with $S_{\mu\nu}$ the axial-gauge  spin~1 polarization sum
\begin{equation}\label{e:defS}
 S_{\mu\nu}:=   \sum_{\lambda=\pm1} \epsilon^{\lambda}_{\mu}(k) (
  \epsilon^{\lambda}_{\nu}(k) )^*
= -\eta_{\mu\nu}+ {(q_{\rm ref})_\mu
    k_\nu+(q_{\rm ref})_\nu k_\mu\over q_{\rm ref}\cdot k}\,,
\end{equation}
where $(q_{\rm ref})_\mu$  is an arbitrary massless reference momentum. That this sum includes only
the two transverse modes can be seen from the condition
\begin{equation}
\eta^{\mu\rho}\eta^{\nu\sigma} \mathcal S_{\mu\nu,\rho\sigma} =2\,,
\end{equation}
corresponding to the normalization condition for the two polarization vectors $\epsilon^{\lambda\lambda}_{\mu\nu}(k)$. Our
work therefore confirms the expected gauge invariance of the quantum correction.

%------------------------------------------------------------------------
\subsection{Comparison with the Feynman graph approach}
\label{sec:comp-with-feynm}
One useful feature of this method is that one can confirm the analysis of
ref.~\cite{BjerrumBohr:2002kt} diagram by diagram. Squaring the tree amplitude
shown in Fig.~\ref{fig:treeemission} leads to discontinuities with the same topology of all the
Feynman diagrams evaluated in~\cite{BjerrumBohr:2002kt}. Evaluating these
individually confirms not only the total result, but also the result of each
of the separate diagrams.\footnote{In~\cite{BjerrumBohr:2002kt} the
  result for each diagrams has been divided by $4m_1m_2$, whereas in this work
the amplitudes are not divided by this factor.} The advantage of doing the diagrams by the unitarity approach is
that one does not have to worry about symmetry factors between Feynman graphs, it is
automatically taken care of by the cut.

The precise relation with the analysis  of~\cite{BjerrumBohr:2002kt}
is the following. We decompose the expression for the tree in~\eqref{e:treeCov1} in
a sum of  three contributions. The first contribution corresponds to the sum of the
graph  in figure~\ref{fig:treeemission}(a)
\begin{equation}\label{e:a}
iM^{(a)}_{\mu\nu,\rho\sigma}(p_1,p_2,-\ell_2,\ell_1)=  \tau_1^{\alpha\beta}(p_1,p_2) \, {i
    \cP_{\alpha\beta;\gamma\delta}\over q^2+i\varepsilon }\,
  \tau_3^{\gamma\delta}{}_{\mu\nu;\rho\sigma}(k_1,k_2,p_1+p_2)\,,
\end{equation}
the second contribution corresponds to the graphs in figure~\ref{fig:treeemission}(b) and~(c) and is given by
\begin{multline}\label{e:bc}
i  M^{(b)+(c)}_{\mu\nu,\rho\sigma}(p_1,p_2,-\ell_2,\ell_1)=  {\tau_{1\,\mu\nu}(p_1,-p_1-\ell_1) \, i \,\tau_{1\,\rho\sigma}(-p_2+\ell_2,p_2)
    \over 2\,p_1\cdot \ell_1+i\varepsilon }\cr
+{\tau_{1\,\rho\sigma}(p_1,-p_1+\ell_2) \, i \,\tau_{1\,\mu\nu}(-p_2+\ell_1,p_2)
    \over -2\,p_1\cdot\ell_2+i\varepsilon }\,.
\end{multline}
The third contribution corresponding to the graph in figure~\ref{fig:treeemission}(d)
\begin{equation}\label{e:d}
iM^{(d)}_{\mu\nu,\rho\sigma}(p_1,p_2,-\ell_2,\ell_1)=\tau_{2\,\mu\nu,\rho\sigma}(p_1,p_2)\,.
\end{equation}
In the cut we get a total of six different contributions from the multiplication of the trees.
Multiplication of the contributions of type~\eqref{e:bc} on both sides of the cut gives the
discontinuity of the box diagram of~\cite[sec.~3.2]{BjerrumBohr:2002kt}. Multiplying the
contribution~\eqref{e:bc} and~\eqref{e:a} leads to the the
discontinuity  of the vertex correction
contributions in figure~5(a) and~5(b) of~\cite[sec.~3.5]{BjerrumBohr:2002kt}. Multiplying the
contribution~\eqref{e:bc} and~\eqref{e:a} leads to the discontinuity
of the triangle
contribution of~\cite[sec.~3.3]{BjerrumBohr:2002kt}. Multiplying the
contribution~\eqref{e:a} and~\eqref{e:a} on both side of the cut gives
the discontinuity of the
vacuum graph contribution in figure~6(a) of~\cite[sec.~3.6]{BjerrumBohr:2002kt}
without the ghost contribution from figure~\ref{fig:ghost}.  Multiplying the
contribution~\eqref{e:a} and~\eqref{e:d} leads to the discontinuity of
the vertex correction
contributions in figure~5(c) and~5(d) of
\cite[sec.~3.5]{BjerrumBohr:2002kt}.
Finally multiplying the
contribution~\eqref{e:d}  on both sides of the cut leads to the
discontinuity  of  the
double-seagull diagrams of~\cite[sec.~3.4]{BjerrumBohr:2002kt}.

%%%%%%%%%%%%%%%%%%%%%%%%%%%%%%%%%%%%%%%%%%%%%%%%%%%%%%%%%%%%%%%%
\section{Matter universality of the quantum corrections}
\label{sec:matter-independence}
%------------------------------------------------------------------------
In this section we will address the previously noted  independence of
the coefficients $C^{NP}$ and $C^{QG}$ on the spin of the external particles. It was found in ref.~\cite{Holstein:2008sx} that the values of
these coefficients are the same for external massive bosons, fermions
or vector external matter. This is what we call matter universality of the coefficients.

Within the unitarity-based methods, the logic for this independence is quite simple.
In a multipole expansion, the on-shell gravitational Compton amplitude has a universal form in the low-energy limit.
For instance the leading contribution in this expansion to the tree-level amplitude---{\it i.e.} the spin-independent contribution considered
in this work---is the same for all types of  external matter fields.
Therefore the discontinuity is  independent in the low-energy limit,
and since we can extract the quantum correction from the
discontinuity, the leading quantum corrections takes the same value
for external massive bosons, massive fermions or massive vector matter.

That universality of the on-shell gravitational Compton amplitude 
can be argued for in various ways. Weinberg~\cite{Weinberg1970} has shown
that the corresponding electromagnetic amplitude is the same for all
external massive bosons, massive fermions or massive vector external matter using only
only gauge invariance. It then follows that the
on-shell gravitational amplitude is also independent of the external
matter type because the
latter can be expressed as the square of the electromagnetic amplitude
as discussed in Section~\ref{sec:treeamp}. Alternatively, as Weinberg
also noted, we know that the electromagnetic amplitude can be
expressed by an effective Lagrangian, whose non-relativistic limit is
determined by the charge and magnetic moment. In the gravitational
case, there is also a low-energy effective Lagrangian for a massive
system, described by its energy-momentum and
spin~\cite{Goldberger:2004jt, Porto:2006bt}.  This yields the leading
couplings of two gravitons to the heavy particle, which is equivalent
to the low-energy limit of our gravitational Compton amplitudes.

In this section we will provide an heuristic general arguments for the
external matter universality  of the
coefficients $C^{NP}$ and $C^{QG}$ based on the KLT amplitude relation.
We will here only consider the spin-independent contribution to the correction of the
classical non-relativistic potential. A more general analysis of the
spin multipole expansion will be done elsewhere.

%------------------------------------------------------------------------
\subsection{The spin~1 case}
\label{sec:spin-1-case}

In the non-relativistic limit the orthogonality conditions  on the
spin~1 polarizations, $p_1\cdot h_1=p_2\cdot h_2=0$, imply that
\begin{equation}
  h_1^0 \simeq {1\over m} \, \vec h_1\cdot \vec p_1, \qquad  h_2^0
  \simeq {1\over m} \, \vec h_2\cdot \vec p_2\,.
\end{equation}
Using the relation $(\vec u\times\vec v)\cdot (\vec x\times\vec y)=(\vec
u\cdot\vec x) (\vec v\cdot \vec y)-(\vec u\cdot y)(\vec v\cdot \vec
x)$ we have the following multipole decomposition
\begin{equation}
  h_1\cdot h_2\simeq -S\, \left(1+ {q^2\over6m^2} \right)   - {i\over 2m^2}\, \vec
  S\cdot (\vec p_1\times \vec p_2) + {1\over m^2} \, \vec p_1\cdot
  \underline Q\cdot \vec p_2\,,
\end{equation}
where $S=\vec h_1\cdot \vec h_2$ is the  spinless singlet, $\vec S:= i \vec h_1\times \vec h_2  $ is the spin vector, and
$\underline Q^{ij}=\frac12\, (h_1^i h_2^j +h_1^j h_2^i)- \frac13
\,\delta^{ij} (\vec h_1\cdot \vec h_2)$ is the (traceless) quadrupole
tensor. We have used that in the non-relativistic limit
$q^2=(p_1+p_2)^2\simeq -2\vec p_1\cdot \vec p_2$.

In the non-relativistic limit we can perform an $1/m$ expansion of the
Compton tree amplitudes.
The Compton scattering of a massive spin~1 vector given in section~\ref{sec:grav-compt-scatt} reads
\begin{eqnarray}\label{e:spin1redux}
  A^{\rm tree}_1(p_1,k_2,p_2,k_1)&=&- (h_1\cdot h_2)\, A^{\rm tree}_0(p_1,k_2,p_2,k_1)\nn \\
\nn
& -& {h_1\cdot F_1\cdot
    F_2\cdot h_2+ (h_1\cdot F_2\cdot h_2)\, (\epsilon_1\cdot p_1)+
    (h_1\cdot F_1\cdot h_2)\, (\epsilon_2\cdot p_2)\over 2 p_1\cdot
    k_1}\nonumber \\
& -& {h_1\cdot F_2\cdot
    F_1\cdot h_2+ (h_1\cdot F_1\cdot h_2)\, (\epsilon_1\cdot p_1)+
    (h_1\cdot F_2\cdot h_2)\, (\epsilon_2\cdot p_2)\over 2 p_1\cdot k_2}\,.
\end{eqnarray}
To leading order in $1/m$ the amplitude approximates to
\begin{eqnarray}\label{e:spin1redux2}
  A^{\rm tree}_1(p_1,k_2,p_2,k_1)&\simeq&   S\, A^{\rm tree}_0(p_1,k_2,p_2,k_1)\nonumber \\
& -& { h_1^i  F_{1\,ij}
    F_2{}^j{}_k  h_2^k+i (\vec S\cdot \vec B_2) \, (\epsilon_1\cdot
    p_1) +i  (\vec S\cdot \vec B_1) \, (\epsilon_2\cdot
    p_2)\over 2 p_1\cdot
    k_1}\nonumber \\
& &- { h_1^i  F_{2\,ij}
    F_1{}^j{}_k  h_2^k+i (\vec S\cdot \vec B_1) \, (\epsilon_1\cdot
    p_1)+i  (\vec S\cdot \vec B_2) \, (\epsilon_2\cdot
    p_2) \over 2 p_1\cdot k_2}\,.
\end{eqnarray}
The first line receives a contribution from the spin-independent
operator $S$ and the last two lines from the spin-orbit and quadrupole
operator. The indices $i, j, k=1,2,3$ run over the spatial components.

The singlet spin-independent contribution $S=\vec h_1\cdot \vec h_2$ in this amplitude is
multiplied by the scalar Compton amplitude. Using the KLT relation the
same property is true for the gravitational Compton
amplitude. Therefore  the spin-independent contribution of the one-loop
correction to Coulomb's potential QED and Newton's potential in
gravity, will be the same as the one finds for scalar scattering,
even with spin~1 external states.

%------------------------------------------------------------------------
\subsection{The spin~$\frac12$ case}
\label{sec:spin-frac12-case}

For the spin~$\frac12$ matter we have a similar decomposition in terms
of a spin-independent piece and a spin-orbit part.
The spin~$\frac12$ amplitude takes the form
\begin{equation}
  A^{\rm tree}_{\frac12}(p_1,k_2,p_2,k_1)= {n_t^{\frac12}\over p_1\cdot k_1}  +
  {n_u^{\frac12}\over p_1\cdot k_2}\,.
\end{equation}
The expression for $n_t^{\frac12}$ is given in eq.~\eqref{e:nt12} with
an equivalent expression for $n_u^{\frac12}$ with the exchange of the
labels $k_1$ and $k_2$.

We start by rewriting these numerators factors using the identity
$(\sla p_1+m)\gamma^\mu u(p_1)= 2 p_1^\mu u(p_1)$, which is a consequence of the
equation of motion $(\sla p_1-m)u(p_1)=0$, to get\footnote{Where we used that
$\{\gamma^\mu,\gamma^\nu\}=2\eta^{\mu\nu}$ and
$\gamma_5=-i\varepsilon_{\mu\nu\rho\sigma} \gamma^{\mu\nu\rho\sigma}$,
and $\gamma^{\mu\nu\rho}=-{i\over3!} \varepsilon^{\mu\nu\rho\sigma} \gamma_5\gamma_\lambda$.}
\begin{equation}
n_t^{\frac12}
=2\bar u(-p_2) \sla \epsilon_2 u(p_1)\,
(\epsilon_1\cdot p_1)-{2\over3}\varepsilon^{\mu\nu\rho\lambda}\,
\epsilon_{2\,\mu} F_{1\,\nu\rho}S_\lambda- 2 \bar u(-p_2)\gamma^\nu u(p_1)
\,  \epsilon_2^\mu F_{1\,\mu\nu}\,.
\end{equation}
Here we have introduced the spin vector
\begin{equation}
  S^\mu := {i\over2}\bar u(-p_2)\gamma_5\gamma^\mu  u(p_1)\,.
\end{equation}
Using Gordon's identities one gets that~\cite{Holstein:2008sw}
\begin{equation}
     \bar u(-p_2)\gamma^\mu u(p_1) ={1\over 1- {q^2\over
         4m^2}}\, \left(S\, {p_1^\mu-p_2^\mu\over2m} + {i\over
       m^2}\varepsilon^{\mu\nu\rho\sigma}  p_{1\,\nu} p_{2\,\rho} S_\sigma\right)\,,
\end{equation}
where $S= \bar u(-p_2) u(p_1)$ is the spinless singlet.

Since our spinors are normalized according to $\bar u(p) u(p)=2m$, following the
conventions of~\cite{Peskin:1995ev}, the non-relativistic limit gives
\begin{eqnarray}
S^\mu&\simeq & -2m\left(0, \vec S:= \frac12\, \xi_2^\dagger \vec \sigma\xi_1\right)\,,  \\
  S&\simeq& -2m\, \left(\xi_2^\dagger \xi_1 +{i\over m^2}   \vec S\cdot
  (\vec p_1\times \vec p_2)\right)\,.
\end{eqnarray}
In this limit,  the numerator factor approximates to
\begin{equation}
   n_t^{\frac12}\simeq (\xi_2^\dagger\xi_1)\, \left(2
(\epsilon_2\cdot p_2)     (\epsilon_1\cdot p_1)  + 2(p_1\cdot k_1)
(\epsilon_1\cdot \epsilon_2)\right)+{2m\over3}
\varepsilon^{\mu\nu\rho} \epsilon_{2\mu} F_{1\nu\rho} S_i\,.
\end{equation}
Therefore the leading $1/m$ expansion of the spin~$\frac12$ Compton
scattering takes the form
\begin{equation}
  \label{eq:3}
  A^{\rm tree}_{\frac12}(p_1,k_2,p_2,k_1)= (\xi_2^\dagger \xi_1) \, A^{\rm tree}_{\rm
    0}(p_1,k_2,p_2,k_1)+{2m\over3}
\varepsilon^{\mu\nu \,i}  S_i\,
\left({ \epsilon_{2\mu} F_{1\nu\rho}\over p_1\cdot k_1}+{ \epsilon_{1\mu} F_{2\nu\rho}\over p_1\cdot k_2} \right)\,.
\end{equation}
We observe that the spin-independent part is again equal to the scalar
amplitude and the spin-orbit part is identical to the one derived for
spin~1 amplitudes.  Using the KLT relation the
same property is true for the gravitational Compton
amplitude. Therefore the spin-independent contribution of the one-loop
correction to Coulomb's potential in QED and in Newton's potential in
gravity, will be the same as the one finds for scalar scattering,
even with massive fermionic external states.

%-------------------------------------------------------------------------
\section{Conclusion}
\label{sec:conclusion}
In this paper we have computed the leading classical and quantum corrections
 to the Coulomb and Newton potentials. This has been done
using modern techniques employing spinor-helicity variables and
on-shell unitarity methods at one-loop order. This is the first time such methods have been applied to compute the quantum correction to the potential.

This approach greatly simplifies the evaluation of these corrections.
It is possible to compare our computation directly to previous Feynman
diagram computations by staying in a covariant formalism, and
explicitly put in the ghost loop contribution. By doing so, we have
verified the gauge invariance of the quantum correction.  Such
unitarity based methods also emphasize that the quantum corrections
come from only the low energy limit of the on-shell gravitational
amplitudes, and are insensitive to the unknown high energy behavior of
the full theory of quantum gravity.

We also considered matter-independence properties of the results
for the non-analytic contributions, and we showed directly using the
KLT formalism that the spinless corrections
to the amplitude theoretically has to be manifestly independent of the nature of the
interacting particles as have been observed in the literature
previously~\cite{Holstein:2008sx}. Such matter-independence statements for
low energy quantum gravity appears to be equivalent to previously noted
statements at low energy in QED~\cite{Weinberg1970}.
The results are low-energy theorems of quantum gravity.

The ultimate and ultraviolet safe theory of quantum gravity is still
not known, however it is gratifying to learn that it is possible to
compute universal results in the theory of quantum gravity. They are universal
in the sense that any theory having the same low-energy spectrum of
particles will have the same answer for the leading corrections independent
of what the high-energy completion might turn out to be.
Although quantum gravity is at times an {\it exhaustive} discipline~\cite{'tHooft:1974bx}
is important to realize that the treament using modern on-shell methods
presents a huge advantage in efficiency.
For example it might be possible to apply some of
our techniques to the recent paper~\cite{Akhoury:2013yua}
and more generally
it might be of interest to reconsider many historical computations in the
light of new computational methods. The recent progress in computational
techniques will here most likely allow an extended analysis.

%-------------------------------------------------------------------------
\section*{Acknowledgements}
We would like to thank Zvi Bern, Poul Damgaard, Simon Badger, Piotr Tourkine,
Barry Holstein and Andreas Ross for discussions.
We  would like to thank Zohar Komargodski for discussions on the matter independence
properties and for pointing out the properties of the non-relativistic
gravitational effective action.
Finally we would like to thank the anonymous referee for comments and
suggestions that helped improving this paper. 
JD and PV would like to thank the Niels Bohr International Academy for
hospitality and financial support, and JD similarly thanks the Institut des Hautes {\'E}tudes Scientifiques.
The research of PV has  been supported by the ANR grant  reference
QFT ANR 12 BS05 003  01, and the CNRS grant PICS number 6076. The research of
JD has been supported by NSF grants PHY-0855119 and PHY-1205986.

%-------------------------------------------------------------------------
\appendix

\section{Vertices and Propagators}\label{sec:vertices}
We will here list the Feynman rules which are employed in our
calculation. For
the derivation of these forms, see~\cite{Donoghue:1993eb,BjerrumBohr:2002ks}.  Our
convention differs from these work by having all incoming momenta.

The propagators are given by
\begin{itemize}
\item The massive scalar propagator is $\displaystyle \frac i{q^2-m^2+i\varepsilon}\,.$
\item The graviton propagator in harmonic gauge can be written in the
  form $\displaystyle \frac {i{\cal
P}^{\alpha\beta,\gamma\delta}}{q^2+i\varepsilon}$
where  $\cP^{\alpha\beta,\gamma\delta}$ is defined in~\eqref{e:defP}.
\end{itemize}
In the background field methods used in
\cite{Donoghue:1993eb,BjerrumBohr:2002ks}, one develops the metric into an expansion
$g_{\mu\nu}=H_{\mu\nu}+\kappa_{(4)} \, h_{\nu\mu}$ where $H_{\mu\nu}$ is
the classical background field and $h_{\mu\nu}$ is the quantum
field. The relation between the vertices given below and the vertices
derived by De Witt is discussed in sec.~\ref{sec:gravcompton}.

The vertices are given by
\begin{itemize}
\item  The 2-scalar-1-graviton vertex $ \tau_1^{\mu
    \nu}(p_1,p_2)$ is
\begin{equation}\label{e:tau1}
\tau_1^{\mu\nu}(p_1,p_2) = \frac{i\kappa_{(4)}}2\left[p_1^\mu p_2^{\nu}
+p_1^\nu p_2^{\mu} -\frac12 \eta^{\mu\nu}\,(p_1+p_2)^2\right]\,.
\end{equation}
\item The 2-scalar-2-graviton vertex $\displaystyle
  \tau_2^{\eta\lambda\rho\sigma}(p_1,p_2)$ is
\begin{eqnarray}\label{e:tau2}
\tau_2^{\eta \lambda \rho \sigma}(p_1,p_2)&=&- {i\kappa_{(4)}^2} \bigg [ \left
\{\cP^{\eta
\lambda, \alpha \delta} \cP^{\rho \sigma,\beta}_{\ \ \ \ \delta} +
\frac14\left\{\eta^{\eta \lambda} \cP^{\rho \sigma,\alpha \beta} +
 \eta^{\rho \sigma} \cP^{\eta \lambda,\alpha \beta} \right \}
\right \} \left (p_{1\,\alpha} p_{2\,\beta} + p_{2\,\alpha} p_{1\,\beta}
\right ) \cr
&+&\frac14 \cP^{\eta \lambda,\rho \sigma} \,( p_1+ p_2)^2\bigg]\,.
\end{eqnarray}
\item The 3-graviton vertex, between two quantum fields and one
classical field, derived via the background field method has the
form~\cite{Donoghue:1993eb}, where $k+q+\pi=0$,
\begin{equation}\label{e:tau3}{
\begin{aligned}
{\tau_3}_{\alpha  \beta \gamma \delta }^{\mu
\nu}(k,q)&=-\frac{i\kappa_{(4)}}2\times
\bigg({\cal P}_{\alpha \beta \gamma \delta }\bigg[k^\mu k^\nu+ \pi^\mu\pi^\nu
+q^\mu q^\nu-
\frac32\eta^{\mu \nu}q^2\bigg]\\[0.00cm]&
+2q_\lambda q_\sigma\bigg[ I_{\alpha \beta }^{\ \ \
\sigma\lambda}I_{\gamma \delta
}^{\ \ \ \mu \nu} + I_{\gamma \delta }^{\ \ \ \sigma\lambda}I_{\alpha
\beta }^{\ \ \
\mu \nu} -I_{\alpha \beta }^{\ \ \ \mu  \sigma} I_{\gamma \delta }^{\ \ \
\nu
\lambda} - I_{\gamma \delta }^{\ \ \ \mu \sigma} I_{\alpha \beta }^{\ \ \
\nu
\lambda}
\bigg]\\[0cm]&
+\bigg[q_\lambda q^\mu \bigg(\eta_{\alpha \beta }I_{\gamma \delta }^{\ \ \
\nu
\lambda}+\eta_{\gamma \delta }I_{\alpha \beta }^{\ \ \ \nu
\lambda}\bigg) +q_\lambda
q^\nu \left(\eta_{\alpha \beta }I_{\gamma \delta }^{\ \ \ \mu
\lambda}+\eta_{\gamma
\delta }I_{\alpha \beta }^{\ \ \ \mu  \lambda}\right)\\&
-q^2\left(\eta_{\alpha
\beta }I_{\gamma \delta }^{\ \ \ \mu \nu}+\eta_{\gamma \delta }I_{\alpha
\beta }^{\
\ \ \mu \nu}\right) -\eta^{\mu \nu}q_\sigma q_\lambda\left(\eta_{\alpha
\beta
}I_{\gamma \delta }^{\ \ \ \sigma\lambda} +\eta_{\gamma \delta }I_{\alpha
\beta }^{\
\ \
\sigma\lambda}\right)\bigg]\\[0cm]&
+\bigg[2q_\lambda\big(I_{\alpha \beta }^{\ \ \ \lambda\sigma}I_{\gamma
\delta
\sigma}^{\ \ \ \ \nu}\pi^\mu +I_{\alpha \beta }^{\ \ \
\lambda\sigma}I_{\gamma
\delta \sigma}^{\ \ \ \ \mu }\pi^\nu +I_{\gamma \delta }^{\ \ \
\lambda\sigma}I_{\alpha \beta \sigma}^{\ \ \ \ \nu}k^\mu +I_{\gamma \delta
}^{\ \ \
\lambda\sigma}I_{\alpha \beta \sigma}^{\ \ \ \ \mu }k^\nu \big)\\&
+q^2\left(I_{\alpha \beta \sigma}^{\ \ \ \ \mu }I_{\gamma \delta }^{\ \ \
\nu
\sigma} + I_{\alpha \beta }^{\ \ \ \nu \sigma}I_{\gamma \delta \sigma}^{\
\ \ \ \mu
}\right) +\eta^{\mu \nu}q_\sigma q_\lambda\left(I_{\alpha \beta }^{\ \ \
\lambda\rho}I_{\gamma \delta  \rho}^{\ \ \ \ \sigma} +I_{\gamma \delta
}^{\ \ \
\lambda\rho}I_{\alpha \beta  \rho}^{\ \ \ \
\sigma}\right)\bigg]\\[0cm]&
+\bigg\{(k^2+\pi^2)\big[\cP_{\alpha \beta }^{\ \ \ \mu  \sigma}\cP_{\gamma
\delta,
\sigma}^{\ \ \ \ \nu} +\cP_{\gamma \delta }^{\ \ \ \mu  \sigma}\cP_{\alpha
\beta,
\sigma}^{\ \ \ \ \nu} -\frac12\eta^{\mu \nu}({\cal P}_{\alpha \beta ,\gamma
\delta
}-\eta_{\alpha\beta}\eta_{\gamma\delta})\big]\\&+\left(\cP_{\gamma \delta }^{\ \ \ \mu \nu}\eta_{\alpha \beta
}\pi^2+\cP_{\alpha
\beta }^{\ \ \ \mu \nu}\eta_{\gamma \delta }k^2\right)\bigg\}\bigg)\,,
\end{aligned}}
\end{equation}
\end{itemize}
where $  I_{\alpha\beta,\gamma\delta}\equiv \cP_{\alpha\beta,\gamma\delta}+
  \frac12\, \eta_{\alpha\beta}\eta_{\gamma\delta}$.
In section~\ref{sec:gravcompton} we explained that the on-shell tree level
amplitudes obtained using these vertices are equivalent to the ones
computed with the vertices given by De Witt~\cite{Dewitt} and
Sannan~\cite{Sannan:1986tz}.  We remark that the expression for
$\tau_3$ is simpler than the three-graviton vertex in these references.

%-------------------------------------------------------------------------
\section{Dispersion relations}\label{sec:dispersion}
In the main text, we calculated the unitarity cut by projecting it onto discontinuities of box, triangle and bubble
integrals. A complementary method involves using the discontinuities to provide the
input to a dispersion relation. We have carried this out in both the
de Donder gauge (with ghosts) and
using the helicity basis (which has only physical degrees of freedom). We briefly
describe the dispersive treatment in this appendix.

The dispersive approach to potentials was pioneered by Feinberg and
Sucher~\cite{Feinberg:1988yw} for QED\footnote{We
have already compared to their QED result in Section~\ref{sec:helicity}.}.
They argue for a dispersive
representation of the scattering potential
\begin{equation}
V(s, q^2) = -\frac{1}{\pi} \int_0^\infty dt \frac{1}{t-q^2} \rho(s,t) {\rm + R.H. cut} \,.
\end{equation}
where the right-hand cut involves only massive states and does not
influence the low energy behavior of the amplitude. Depending on the
ultimate high energy theory, this dispersion relation may require
subtractions. However, an important point is that the subtraction
constants are analytic functions of powers of $q^2$. The subtraction
constants then are related to local, analytic terms in the effective
Lagrangian~\cite{Donoghue:1996kw}, and cannot modify the non-analytic
terms that come from the low energy end of the dispersion relation. In
the case of gravity, the subtraction constants correspond to higher
curvature terms in the gravitational action. If we are interested in
the low-energy non-analytic terms we can use either subtracted or
unsubtracted forms of the dispersion relation.

The spectral function $\rho(s,t)$ is formed by multiplying together the on-shell gravitational Compton amplitudes.
In the axial gauge of the helicity basis we have only the physical degrees of freedom
\begin{equation} \label{e:rhohelicity}
  \rho (s,t)=-{1\over \pi}\int {d\Omega_\ell\over 4\pi }\,
 M^{\rm
    tree}_{\mu\nu,\rho\sigma}(p_1,p_2,-\ell_2,\ell_1)
  \mathcal S^{\mu\nu,\alpha\beta} \mathcal S^{\rho\sigma,\gamma\delta} (M^{\rm
    tree}_{\alpha\beta,\gamma\delta}(p_4,p_3,\ell_2,-\ell_1))^* \,,
\end{equation}
where $\mathcal S_{\mu\nu,\rho\sigma}$ is the polarization sum of Eq. \ref{e:defSS}. The graviton momenta in the numerator are
taken to be on-shell. If we work in
harmonic gauge we have a similar relation with the harmonic gauge polarization sum of Eq. \ref{e:defP}
\begin{equation}
  \label{e:rhoharmonic}
 \rho (s,t)=-{1\over \pi}\int {d\Omega_\ell\over 4\pi}\,
 M^{\rm
    tree}_{\mu\nu,\rho\sigma}(p_1,p_2,-\ell_2,\ell_1)
  \cP^{\mu\nu,\alpha\beta} \cP^{\rho\sigma,\gamma\delta} (M^{\rm
    tree}_{\alpha\beta,\gamma\delta}(p_4,p_3,\ell_2,-\ell_1))^*
  \ .
\end{equation}
Of course, in the harmonic gauge we expect to also need to include ghost fields, and this will be verified.

Feinberg and Sucher describe how to do the angular phase-space integrals. It is useful to
go to the frame where $p_1 = (\omega, {\vec p})$, $p_2 = (\omega, -{\vec p})$ with ${\vec p} = im_1\zeta_1 {\hat p}$ and $\zeta_1=\sqrt{1-t/4m_1^2}$.
In the gravity case there are more momentum factors in the numerator than with QED, but the phase space
integrals are simple generalizations of the ones described in~\cite{Feinberg:1988yw}.
After the phase-space integration, the spectral functions can be expanded at low-energy with the form
\begin{equation}
\rho (s,t) = a_1(s)\frac{1}{\sqrt{t}} + a_2(s) + ....
\end{equation}
yielding a potential function
\begin{equation}
V(s,q^2) ={1\over \pi}[ a_1(s)\frac{\pi}{\sqrt{-q^2}} + a_2(s) \ln(-q^2) + .... ]
\end{equation}
which is to be evaluated in the non-relativistic limit.

We have carried out this program in both the helicity basis and in
harmonic gauge. In the helicity basis, for simplicity we chose the
reference momentum for $\ell_1$ to be $\ell_2$ and visa-versa. The
covariant amplitudes were multiplied together, the phase-space
integral done and the result was Taylor expanded at low energy using
Mathematica.  In the helicity basis, this directly reproduced both the
classical and quantum non-analytic terms as described in the text.
For the harmonic gauge calculation, ghost fields were needed and a
separate spectral function for ghosts was included, with the sum of
graviton and ghost effects again yielding the expected answer.

The main technical difference between the methods described in the
text and this dispersive method is that in the latter method the phase
space integral is explicitly calculated while in the former the
discontinuity is used to identify the contributions of box, triangle
and bubble diagrams. Of course, these yield the same results because
the box, triangle and bubble diagrams respect the causality and
analyticity properties that go into the dispersion relations.

%%%%%%%%%%%%%%%%%%%%%%%%%%%%%%%%%%%%%%%%%%%%%%%%%%%%%%%%%%%%%%%%%%

\end{document}